\documentclass[fleqn,usenatbib]{mnras}

\usepackage{newtxtext,newtxmath}

\usepackage[T1]{fontenc}

\DeclareRobustCommand{\VAN}[3]{#2}
\let\VANthebibliography\thebibliography
\def\thebibliography{\DeclareRobustCommand{\VAN}[3]{##3}\VANthebibliography}

\usepackage{graphicx}
\usepackage{multirow}
\usepackage{graphicx}	
\usepackage{amsmath}	
\usepackage{placeins}
\usepackage{xfrac}
\usepackage{mathtools}

\usepackage{hyperref}
\hypersetup{colorlinks=true, linkcolor=Navy, citecolor=Navy, urlcolor=Navy}

\usepackage{color}

\definecolor{DarkRed}           {RGB}{139,   0,   0}
\definecolor{Red}               {RGB}{255,   0,   0}
\definecolor{Firebrick}         {RGB}{178,  34,  34}
\definecolor{Crimson}           {RGB}{220,  20,  60}
\definecolor{IndianRed}         {RGB}{205,  92,  92}
\definecolor{LightCoral}        {RGB}{240, 128, 128}
\definecolor{Salmon}            {RGB}{250, 128, 114}
\definecolor{DarkSalmon}        {RGB}{233, 150, 122}
\definecolor{LightSalmon}       {RGB}{255, 160, 122}

\definecolor{OrangeRed}         {RGB}{255,  69,   0}
\definecolor{Tomato}            {RGB}{255,  99,  71}
\definecolor{DarkOrange}        {RGB}{255, 140,   0}
\definecolor{Coral}             {RGB}{255, 127,  80}
\definecolor{Orange}            {RGB}{255, 165,   0}

\definecolor{DarkKhaki}         {RGB}{189, 183, 107}
\definecolor{Gold}              {RGB}{255, 215,   0}
\definecolor{Khaki}             {RGB}{240, 230, 140}
\definecolor{PeachPuff}         {RGB}{255, 218, 185}
\definecolor{Yellow}            {RGB}{255, 255,   0}
\definecolor{PaleGoldenrod}     {RGB}{238, 232, 170}
\definecolor{Moccasin}          {RGB}{255, 228, 181}
\definecolor{PapayaWhip}        {RGB}{255, 239, 213}
\definecolor{LightGoldenrodYellow}{RGB}{250, 250, 210}
\definecolor{LemonChiffon}      {RGB}{255, 250, 205}
\definecolor{LightYellow}       {RGB}{255, 255, 224}

\definecolor{Maroon}            {RGB}{128,   0,   0}
\definecolor{Brown}             {RGB}{165,  42,  42}
\definecolor{SaddleBrown}       {RGB}{139,  69,  19}
\definecolor{Sienna}            {RGB}{160,  82,  45}
\definecolor{Chocolate}         {RGB}{210, 105,  30}
\definecolor{DarkGoldenrod}     {RGB}{184, 134,  11}
\definecolor{Peru}              {RGB}{205, 133,  63}
\definecolor{RosyBrown}         {RGB}{188, 143, 143}
\definecolor{Goldenrod}         {RGB}{218, 165,  32}
\definecolor{SandyBrown}        {RGB}{244, 164,  96}
\definecolor{Tan}               {RGB}{210, 180, 140}
\definecolor{Burlywood}         {RGB}{222, 184, 135}
\definecolor{Wheat}             {RGB}{245, 222, 179}
\definecolor{NavajoWhite}       {RGB}{255, 222, 173}
\definecolor{Bisque}            {RGB}{255, 228, 196}
\definecolor{BlanchedAlmond}    {RGB}{255, 235, 205}
\definecolor{Cornsilk}          {RGB}{255, 248, 220}

\definecolor{DarkGreen}         {RGB}{  0, 100,   0}
\definecolor{Green}             {RGB}{  0, 128,   0}
\definecolor{DarkOliveGreen}    {RGB}{ 85, 107,  47}
\definecolor{ForestGreen}       {RGB}{ 34, 139,  34}
\definecolor{SeaGreen}          {RGB}{ 46, 139,  87}
\definecolor{Olive}             {RGB}{128, 128,   0}
\definecolor{OliveDrab}         {RGB}{107, 142,  35}
\definecolor{MediumSeaGreen}    {RGB}{ 60, 179, 113}
\definecolor{LimeGreen}         {RGB}{ 50, 205,  50}
\definecolor{Lime}              {RGB}{  0, 255,   0}
\definecolor{SpringGreen}       {RGB}{  0, 255, 127}
\definecolor{MediumSpringGreen} {RGB}{  0, 250, 154}
\definecolor{DarkSeaGreen}      {RGB}{143, 188, 143}
\definecolor{MediumAquamarine}  {RGB}{102, 205, 170}
\definecolor{YellowGreen}       {RGB}{154, 205,  50}
\definecolor{LawnGreen}         {RGB}{124, 252,   0}
\definecolor{Chartreuse}        {RGB}{127, 255,   0}
\definecolor{LightGreen}        {RGB}{144, 238, 144}
\definecolor{GreenYellow}       {RGB}{173, 255,  47}
\definecolor{PaleGreen}         {RGB}{152, 251, 152}

\definecolor{Teal}              {RGB}{  0, 128, 128}
\definecolor{DarkCyan}          {RGB}{  0, 139, 139}
\definecolor{LightSeaGreen}     {RGB}{ 32, 178, 170}
\definecolor{CadetBlue}         {RGB}{ 95, 158, 160}
\definecolor{DarkTurquoise}     {RGB}{  0, 206, 209}
\definecolor{MediumTurquoise}   {RGB}{ 72, 209, 204}
\definecolor{Turquoise}         {RGB}{ 64, 224, 208}
\definecolor{Aqua}              {RGB}{  0, 255, 255}
\definecolor{Cyan}              {RGB}{  0, 255, 255}
\definecolor{Aquamarine}        {RGB}{127, 255, 212}
\definecolor{PaleTurquoise}     {RGB}{175, 238, 238}
\definecolor{LightCyan}         {RGB}{224, 255, 255}

\definecolor{Navy}              {RGB}{  0,   0, 128}
\definecolor{DarkBlue}          {RGB}{  0,   0, 139}
\definecolor{MediumBlue}        {RGB}{  0,   0, 205}
\definecolor{Blue}              {RGB}{  0,   0, 255}
\definecolor{MidnightBlue}      {RGB}{ 25,  25, 112}
\definecolor{RoyalBlue}         {RGB}{ 65, 105, 225}
\definecolor{SteelBlue}         {RGB}{ 70, 130, 180}
\definecolor{DodgerBlue}        {RGB}{ 30, 144, 255}
\definecolor{DeepSkyBlue}       {RGB}{  0, 191, 255}
\definecolor{CornflowerBlue}    {RGB}{100, 149, 237}
\definecolor{SkyBlue}           {RGB}{135, 206, 235}
\definecolor{LightSkyBlue}      {RGB}{135, 206, 250}
\definecolor{LightSteelBlue}    {RGB}{176, 196, 222}
\definecolor{LightBlue}         {RGB}{173, 216, 230}
\definecolor{PowderBlue}        {RGB}{176, 224, 230}

\definecolor{Indigo}            {RGB}{ 75,   0, 130}
\definecolor{Purple}            {RGB}{128,   0, 128}
\definecolor{DarkMagenta}       {RGB}{139,   0, 139}
\definecolor{DarkViolet}        {RGB}{148,   0, 211}
\definecolor{DarkSlateBlue}     {RGB}{ 72,  61, 139}
\definecolor{BlueViolet}        {RGB}{138,  43, 226}
\definecolor{DarkOrchid}        {RGB}{153,  50, 204}
\definecolor{Fuchsia}           {RGB}{255,   0, 255}
\definecolor{Magenta}           {RGB}{255,   0, 255}
\definecolor{SlateBlue}         {RGB}{106,  90, 205}
\definecolor{MediumSlateBlue}   {RGB}{123, 104, 238}
\definecolor{MediumOrchid}      {RGB}{186,  85, 211}
\definecolor{MediumPurple}      {RGB}{147, 112, 219}

\definecolor{MistyRose}         {RGB}{255, 228, 225}
\definecolor{AntiqueWhite}      {RGB}{250, 235, 215}
\definecolor{Linen}             {RGB}{250, 240, 230}
\definecolor{Beige}             {RGB}{245, 245, 220}
\definecolor{WhiteSmoke}        {RGB}{245, 245, 245}
\definecolor{LavenderBlush}     {RGB}{255, 240, 245}
\definecolor{OldLace}           {RGB}{253, 245, 230}
\definecolor{AliceBlue}         {RGB}{240, 248, 255}
\definecolor{Seashell}          {RGB}{255, 245, 238}
\definecolor{GhostWhite}        {RGB}{248, 248, 255}
\definecolor{Honeydew}          {RGB}{240, 255, 240}
\definecolor{FloralWhite}       {RGB}{255, 250, 240}
\definecolor{Azure}             {RGB}{240, 255, 255}
\definecolor{MintCream}         {RGB}{245, 255, 250}
\definecolor{Snow}              {RGB}{255, 250, 250}
\definecolor{Ivory}             {RGB}{255, 255, 240}
\definecolor{White}             {RGB}{255, 255, 255}

\definecolor{Black}             {RGB}{  0,   0,   0}
\definecolor{DarkSlateGray}     {RGB}{ 47,  79,  79}
\definecolor{DimGray}           {RGB}{105, 105, 105}
\definecolor{SlateGray}         {RGB}{112, 128, 144}
\definecolor{Gray}              {RGB}{128, 128, 128}
\definecolor{LightSlateGray}    {RGB}{119, 136, 153}
\definecolor{DarkGray}          {RGB}{169, 169, 169}
\definecolor{Silver}            {RGB}{192, 192, 192}
\definecolor{LightGray}         {RGB}{211, 211, 211}
\definecolor{Gainsboro}         {RGB}{220, 220, 220}

\definecolor{DavysGray}         {RGB}{ 85,  85,  85}
\definecolor{Jet}               {RGB}{ 52,  52,  52}

\newcommand{\AAA}{\textup{\AA}}

\newcommand{\kmps}{\textup{km}\,\textup{s}^{-1}}
\newcommand{\mps}{\textup{m}\,\textup{s}^{-1}}
\newcommand{\cmps}{\textup{cm}\,\textup{s}^{-1}}

\newcommand{\VAL}[1]{{{#1}}}

\newcommand{\Color}[1]{\color{black}}

\title[BLUVES II]{Test of a 34\:GHz EOM laser frequency comb at ESPRESSO}

\author[Tobias M. Schmidt et al.]{%
Tobias M. Schmidt,$^{1,2}$\thanks{E-mail: tobias.schmidt2@uni-goettingen.de}
Ewelina Obrzud,$^{3}$ 
Fran\c{c}ois Bouchy$^{1}$, 
Gaspare Lo Curto$^{4}$, 
Victor Brasch$^{5}$, 
\newauthor Tobias Herr$^{6,7}$, 
Furkan Ayhan$^{8}$, 
Severine Denis$^{3}$, 
Davide Grassani$^{3}$, 
Jean Berney$^{3}$, 
Bruno Chazelas$^{1}$, 
\newauthor Weichen Fan$^{6}$, 
Jannis Holzer$^{3}$, 
Ian Hughes$^{1}$, 
Markus Ludwig$^{9}$, 
Antonio Manescau$^{10}$, 
Luca Pasquini$^{10}$, 
\newauthor Francesco Pepe$^{1}$, 
Luis Guillermo Villanueva$^{8}$, 
Fran\c{c}ois Wildi$^{1}$, 
Thibault Wildi$^{6}$ 
\\
  $^{1}$Observatoire Astronomique de l’Universit\'e de Gen\`eve, Chemin Pegasi 51, Sauverny, CH-1290, Switzerland\\
  $^{2}$Institut für Astrophysik und Geophysik, Georg-August-Universität, 37077 Göttingen, Germany\\
  $^{3}$Swiss Center for Electronics and Microtechnology (CSEM), Rue de l'Observatoire 58, Neuchatel, CH-2002, Switzerland\\
  $^{4}$European Southern Observatory (ESO), Av. Alonso de Cordova 3107,  Casilla 19001, Santiago de Chile, Chile\\
  $^{5}$Q.ANT GmbH, 70565 Stuttgart, Germany\\
  $^{6}$Deutsches Elektronen-Synchrotron DESY, Notkestr. 85, 22607 Hamburg, Germany\\
  $^{7}$Department of Phyiscs, Universität Hamburg UHH, Luruper Chaussee 149, 22607 Hamburg, Germany\\
  $^{8}$École Polytechnique Fédérale de Lausanne (EPFL), 1015 Lausanne, Switzerland\\
  $^{9}$Deutsches Elektronen-Synchrotron DESY, Notkestr. 85, 22607 Hamburg, Germany\\
  Present address: University of Luxembourg, 162a, Avenue de la Faïencerie, L-1511 Luxembourg, Luxembourg, and\\
  Institute for Advanced Studies, University of Luxembourg, Campus Belval, L-4365 Esch-sur-lzette, Luxembourg\\
  $^{10}$European Southern Observatory, Karl-Schwarzschild-Strasse 2, 85748, Garching b. München, Germany
}

\date{Accepted XXX. Received YYY; in original form ZZZ}

\pubyear{2025}

\newcommand\blfootnote[1]{%
  \begingroup
  \renewcommand\thefootnote{}\footnote{#1}%
  \addtocounter{footnote}{-1}%
  \endgroup
}

\begin{document}
\label{firstpage}
\pagerange{\pageref{firstpage}--\pageref{lastpage}}
\maketitle

\begin{abstract}
Laser frequency combs (LFCs) are a promising technology for wavelength calibration of astronomical high-resolution spectrographs requiring utmost accuracy and stability, since they directly translate the fundamental SI time standard from the radio frequency regime to optical frequencies.
However, they have so far seen limited use in practice, due to their complexity, incomplete wavelength coverage, but also the challenges in the data analysis they imply.
Here, we present a detailed test of a \VAL{$34\,\mathrm{GHz}$} electro-optic modulation comb with the ESPRESSO spectrograph. Using thin-film lithum-niobate waveguides for broadening and harmonic generation, the setup provides partial coverage of the IR, visible, and near-UV spectral ranges.
We focus on assessing the quality of the delivered spectra and their capability to facilitate accurate and stable wavelength calibration. We present a detailed analysis of the spectrally-diffuse background, the line width, and characterize the line-spread function over a broader width than possible with the ESPRESSO facility LFC.
Comparing both combs, we find strong local discrepancies in the wavelength calibration accuracy up to \VAL{$15\,\mps$}, which correlate with the echellogram structure.
These do not originate from the lasers, but from misalignments in the ESPRESSO calibration unit, highlighting the strong need to make instrument fiber feeds more robust to light-injection effects.
Nevertheless, we demonstrate excellent stability of the wavelength calibration, with a scatter of only \VAL{$17\,\cmps$}. This, however, can only be achieved when accurately modeling the non-Gaussian line-spread function, showcasing the need for advanced data analysis techniques when dealing with LFC spectra.
\end{abstract}



\section{Introduction}

\blfootnote{Based on observations collected at the European Southern Observatory under ESO program 114.28HD.001.}%
Numerous astrophysical science cases require extremely \textit{precise}, \textit{accurate}, and \textit{stable} wavelength calibration of the spectrograph%
\footnote{Throughout this paper, the terms \textit{precision}, \textit{accuracy}, and \textit{stability} are used as defined in \citet{Martins2024} and described more extensively in the appendix of \citet{Schmidt2024}.}.
Well known here is the hunt for extrasolar planets using the radial-velocity method (RV, \citealt{Griffin1967, Baranne1979, Mayor1983,Mayor1995}), but also the search for a variation of fundamental physical constants \citep[e.g.][]{Webb1999,Martins2017,Murphy2021} requires a high-fidelity and here in particular very \textit{accurate} wavelength calibration.
Above all, the redshift drift experiment \citep{Sandage1962,Liske2008, Cristiani2023}, one of the main science drivers of the ArmazoNes high Dispersion Echelle Spectrograph \citep[ANDES,][]{Marconi2022, Marconi2024, Martins2024}, the future high-resolution spectrograph for the European Extremely Large Telescope, imposes extreme and unprecedented requirements on the \textit{stability} of the wavelength calibration over timescales of decades, to be able to unambiguously detect the cosmological expansion of the Universe in real time.

Accurate and stable wavelength calibration of a spectrograph requires a calibration source that provides a large number of spectral features with precisely known wavelengths. One promising technology for this are laser frequency combs \citep[LFC, e.g.][]{Udem2002, McCracken2017a, Fortier2019, Herr2019, Diddams2020}.
Their characteristic feature is that they provide a large number of laser lines whose frequencies follow a well defined relation of the form
\begin{equation}
  f_{i} = f_\mathrm{0} + i \cdot f_\mathrm{rep}.
 \label{Eq:Comb}
\end{equation}
Here, $i$ is just an integer counting the mode number, while $f_\mathrm{0}$ and $f_\mathrm{rep}$ are the comb-defining frequencies, namely the offset frequency or carrier-envelope offset frequency, and the repetition rate defining the separation of individual lines in the spectrum.
Both of these frequencies are typically in the MHz to GHz range and can be easily measured electronically and actively stabilized \citep{Reichert1999, Telle1999,Jones2000}, for instance against a global navigation satellite system (GNSS) disciplined atomic clock.
LFCs therefore allow to accurately phase-coherently link frequency information between the radio and optical regime.

To calibrate an astronomical spectrograph with a LFC, one requires a line separation that can be fully resolved. Even for high-resolution echelle spectrographs with an resolving power of $R=\frac{\lambda}{\Delta\lambda}\approx100\,000$, this requires\,--\,depending on the wavelength range\,--\,repetition rates of the order of tens of GHz.
Fiber-based mode-locked LFCs, however, typically operate at $80$ to $250\,\mathrm{MHz}$ and require complex filtering cavities to suppress unwanted modes and thereby increase the separation of the transmitted lines \citep[e.g.][]{Steinmetz2008, Wilken2010a, Wilken2012, Ycas2012, Probst2014,Probst2016, Nakamura2023, Wu2024, Tian2024}.
The situation is similar for Ti:Sapphire-based systems, which typically run with repetition rates around $1\,\mathrm{GHz}$ \citep[e.g.][]{Phillips2012a, Depagne2016, McCracken2017b, Chae2021, Cheng2024, Newman2025}.
An alternative are electro-optic modulation combs (EOM) which directly create a large mode spacing by phase- and amplitude modulation of a continuous-wave (CW) laser \citep[e.g.][]{Yi2016, Kokubo2016, Obrzud2018, Obrzud2019, Metcalf2019a, Metcalf2019b, Ludwig2024, Sekhar2024}, or microresonators \citep{Suh2019, Obrzud2019a, Ludwig2024}.

Another challenge for LFCs is to achieve the large wavelength coverage of astronomical spectrographs, in particular in the blue and near-UV spectral range. Since there are no suitable laser-gain media in the visible regime, all combs developed for astronomical spectrograph calibration are initially defined over relatively narrow spectral ranges at infrared wavelengths, around $800\,\mathrm{nm}$ (Ti:Sapphire), $1\,\mathrm{\mu{}m}$ (Ytterbium doped fibers), or $1.5\,\mathrm{\mu{}m}$ (Erbium doped fibers). These then have to be spectrally broadened and frequency-converted into the optical regime using non-linear processes in suitable media and at very high power densities. Possible solutions for this are the use of highly non-linear fibers, (tapered) photonic crystal fibers \citep[e.g][]{Stark2010, Wilken2012}, bulk material like e.g. barium boron oxide, BBO, \citep[e.g.][]{Metcalf2019a}, or waveguides in e.g. silicon nitride \citep[e.g.][]{Obrzud2019} or lithium niobate optical chips \citep[e.g.][]{Wu2024, Ludwig2024}.

In \citet{Ludwig2024}, we have presented an $18\,\mathrm{GHz}$ EOM comb, combined with spectral broadening in nanophotonic lithium niobate waveguides, and demonstrated wavelength calibration in the visible and near-UV for the SOPHIE spectrograph at Observatoire de Haute Provence (OHP/SOPHIE). 
Lithium niobate provides strong $\chi^{(3)}$ and $\chi^{(2)}$ non-linearities, and periodic poling allows engineering of the the phase-matching conditions for efficient, broadband frequency conversion. Our thin-film lithium niobate (TFLN) waveguides thus facilitate efficient spectral broadening and harmonic generation \citep{Ayhan2025}. Starting from a seed laser at \VAL{$1.56\,\mathrm{\mu{}m}$}, the third harmonic of the fundamental comb then covers a spectral domain around \VAL{$520\,\mathrm{nm}$} and the fourth one around \VAL{$390\,\mathrm{nm}$}, allowing unprecedented coverage of the near-UV region. However, our test at OHP also demonstrated the impact  phase noise in the optical pulse can have on the spectrally-diffuse background in the LFC spectrum, a serious limitation for using LFCs in practice to calibrate astronomical spectrographs.

Here, we thus present an improved setup, focusing explicitly on the suppression of modal noise and background flux, and tested it with ESPRESSO, the Echelle SPectrograph for Rocky Exoplanets and Stable Spectroscopic Observations \citep{Pepe2021}. The superior capabilities of ESPRESSO allow us to characterize the LFC spectra in much greater detail than possible with OHP/SOPHIE and to test the accuracy and stability of the derived wavelength calibration at an exquisite level.

\section{Instrumental Setup and Observations}
\label{Sec:Data}

ESPRESSO is the current flagship for stable and accurate astronomical spectroscopy \citep{Megevand2014, GonzalezHernandez2018, Pepe2021}.
It is fiber-fed, offers a resolving power of $R = \frac{\lambda}{\Delta{}\lambda}\approx 135\,000$ in the standard \texttt{HR} observing mode, continuous wavelength coverage from \VAL{$3785$} to \VAL{$7885\,\AAA$}, and was specifically designed for RV surveys of rocky exoplanets and tests of fundamental physics. Special emphasis was therefore put in instrument stability and the ability to derive accurate wavelength calibrations.
ESPRESSO is equipped with a full set calibration sources, in particular a thorium-argon hollow-cathode lamp (ThAr HCL) and a white-light passively-stabilized Fabry-P\'erot etalon (FP) with a free spectral range (FSR) of \VAL{$\approx19.7\,\mathrm{GHz}$} \citep[e.g.][]{Wildi2011}. These provide the default wavelength calibration and are used for almost all science programs. In addition, the suite of calibration sources encompasses a $18\,\mathrm{GHz}$ mode-locked LFC
, similar to the one described in \citet{Probst2014,Probst2016}, and thus allows detailed comparisons and tests between sources.
Over the years, significant efforts have been invested to characterize, improve, and validate the ESPRESSO wavelength calibration accuracy and stability \citep[e.g.][]{Schmidt2021, Schmidt2022, Schmidt2024, Schmidt2025}.
%
ESPRESSO is therefore the ideal instrument to test our $34\,\mathrm{GHz}$ EOM comb under realistic conditions.

We therefore conducted in December 2024 a dedicated mission, spanning approximately two weeks, to the Paranal observatory and tested our comb there with ESPRESSO.
Significant time was required to unpack and integrate the complex setup at the mountain, as well as to disassemble everything again and prepare for shipment back to Europe. This left about one week, from December 6 to 13, for detailed tests and experiments with the laser setup.

ESPRESSO can be fed by two fibers simultaneously, dedicated to science target and sky, or two calibration sources. In addition, it employs a pupil slicer in its design \citep{dellAgostino2014} that images the spectrum of each fiber twice onto the detector. The spectra from each input fibers and slice are extracted separately and can be used for cross-checks.
All spectra were taken in \texttt{1HR2x1} mode, i.e. at the standard resolving power of $R\approx135\,000$ and with 2$\times$ binning in cross-dispersion and 1$\times$ in spectral direction.

All data was acquired during daytime and no nighttime observations were conducted, since the goal of the test was to demonstrate the ability of our comb setup to calibrate astronomical spectrographs and to compare wavelength information derived from different kinds of sources. Observations of celestial objects are not ideally suited for this, because e.g. stars are inherently noisy at the level of $\approx0.5\,\mps$ \citep[e.g.][]{Gonzalez2024,Hobson2024,Figueira2025} and require complex modeling of stellar activity to find or verify small RV signals. A true benchmark of an instrument or light source is therefore better conducted against other calibration sources, which can be done during daytime without requiring precious nighttime observations.

\label{Sec:DataReduction}

Data processing starting from the raw frames, in particular spectral extraction and wavelength calibration, for all spectra presented here was done with a custom data-reduction pipeline \citep{Schmidt2021,Schmidt2024}. Spectral extraction was performed using a variant of the \textit{flat-relative optimal extraction} algorithm presented by \citet{Zechmeister2014}, with some special modifications to provide a highly accurate removal of scattered light.

The wavelength calibration of the spectra is based either on the combined ThAr/FP wavelength solution \citep{Bauer2015, Cersullo2019, Schmidt2021} or the independent LFC solution, and derived from the standard calibration exposures taken every morning. Due to technical issues, usable exposures of the $18\,\mathrm{GHz}$ facility LFC are only available for December 5, 10, and 13. Due to the excellent stability of ESPRESSO and the simultaneous drift measurement, these calibrations are nevertheless applicable to all our datasets.

Crucial for an accurate and stable wavelength calibration but also for the general analysis of the spectra is a careful modeling of the instrumental line-spread function (LSF).
This is done in a fully identical way as described by \citet{Schmidt2024}.
From the spectra of the $18\,\mathrm{GHz}$ LFC, which provide a large number of truly unresolved lines and therefore reveal the instrumental profile of the spectrograph, a non-parametric model of the LSF is derived. This is done separately for each spectrograph fiber and slice, and in 16 blocks per echelle order. These accurate models of the ESPRESSO LSF are then employed for every line fit, except explicitly stated.

\section{Laser Setup}
\label{Sec:LaserSetup}

\begin{figure*}
 \includegraphics[width=\linewidth]{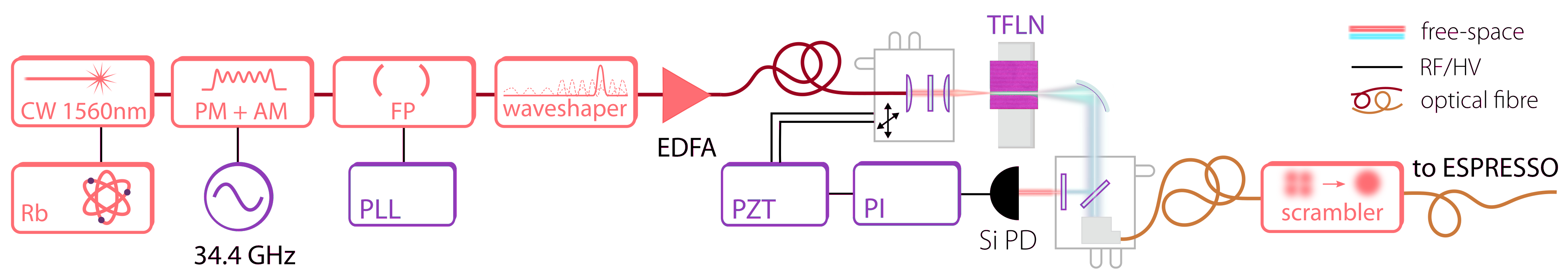}
 \caption{ The LFC scheme. Ultrashort optical pulses are generated via electro-optic modulation and delivered in a polarisation maintaining optical fibers to the TFLN  waveguide alignment setup. Light from the fiber is collimated, polarization aligned and focused onto the input facet of the waveguide. At the waveguide output, the generated broadband UV-VIS spectrum is collimated with an off axis parabolic mirror and then injected into the output multimode fiber. In front of the fiber, part of the light is bandpass-filtered around \VAL{$700\,\mathrm{nm}$}  to provide the signal for automatic alignment of the input beam to the waveguide via two piezo actuators. Before entering the ESPRESSO spectrograph, the light is mode-scrambled. CW – continuous wave, PM – phase modulation, AM – amplitude modulation, FP – Fabry-Perot, EDFA – erbium-doped fibre amplifier, TFLN - thin film lithium nobiate, PLL – phase-locked loop, PZT – piezo transducer, PI – proportional-integral, Si PD – silicon photodiode, RF – radio frequency, HV – high voltage
 }
 \label{Fig:LFCsetup}
\end{figure*}

The LFC in our setup is based on electro-optic modulation and operates with a repetition rate of \VAL{$34.4\,\mathrm{GHz}$}, with a laser center wavelength of \VAL{$1560\,\mathrm{nm}$}. Figure~\ref{Fig:LFCsetup} shows the scheme of the LFC system. The continuous-wave (CW) seed laser is modulated in phase and amplitude with the modulation frequency equal to the intended repetition rate of \VAL{$34.4\,\mathrm{GHz}$}. The phase modulation imprints a frequency chirp onto the CW light, while the amplitude modulation carves out the right sign of chirp for the following optical pulse synthesis. To ensure accuracy and stability of the laser comb, the CW seed laser is stabilized to a rubidium atomic frequency standard. Specifically, the second harmonic of the laser is locked via the Pound-Drever-Hall (PDH) technique to the \VAL{Rb D$_2$ CO: 22--23} hyperfine transition; a field programmable gate array (FPGA) is used to implement the lock.

As the electro-optic modulation results in a rather high phase noise added to the carrier, an optical Fabry-P\'erot cavity (FP) with the free spectral range (FSR) equal to the repetition rate of the LFC is employed to filter the high frequency phase noise. The cavity is characterized by a finesse of \VAL{725} and is stabilized to the seed laser using the PDH locking scheme. Once the cavity is stabilized, the repetition rate of the LFC is adjusted to the cavity's free spectral range (FSR) by maximizing its optical transmission. 

Next, the optical pulses are synthesized and shaped by applying the right amount of dispersion with a waveshaper and subsequently amplified up to \VAL{$4.5\,\mathrm{W}$} 
and sent through a carefully tailored sequence of highly nonlinear optical fibers for optical pulse compression down to \VAL{$\sim50\,\mathrm{fs}$} temporal width.
The entire system up to this point, but excluding the Fabry-P\'erot cavity, is based on polarization-maintaining fibers.

\begin{figure}
 \includegraphics[width=\linewidth]{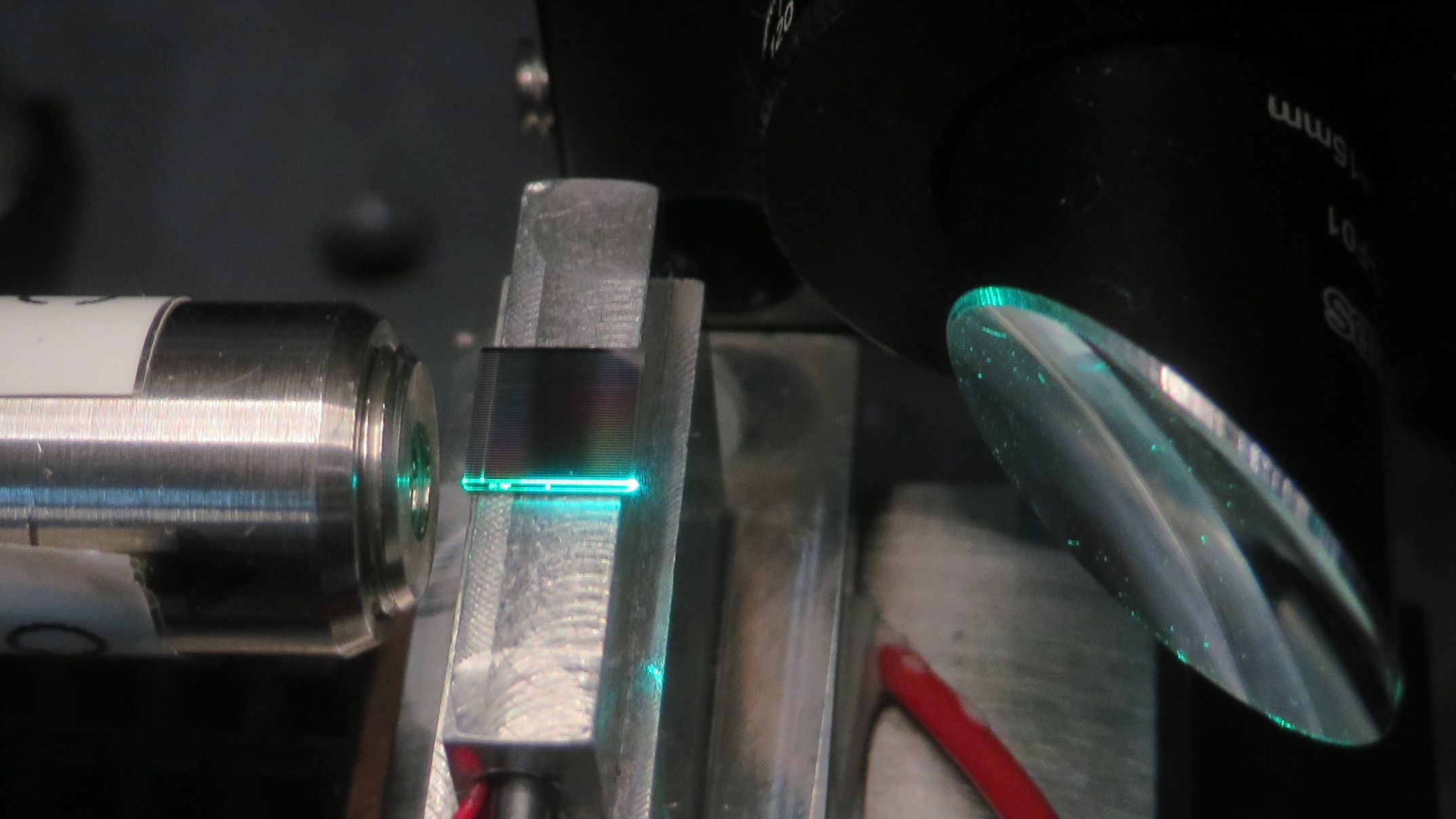}
 \caption{Optical setup used for spectral broadening and harmonic generation. A lens (left) focuses the high-power IR light and injects it into the waveguide on the TFLN optical chip (center) where it is converted into the visible regime. At the output, the light is collected by a a parabolic mirror (right).}
 \label{Fig:OpticalChip}
\end{figure}

The frequency conversion to reach the UV/VIS wavelength range is performed in a highly-nonlinear, periodically-poled, TFLN integrated waveguide. Cascaded, phase-matched second- and higher-order nonlinear processes facilitate the transfer of the IR spectra to the UV/VIS wavelengths via harmonic generation. Given that the input IR spectrum is reasonably broadband, and the chromatic dispersion at the UV/VIS wavelengths are strong, we use an aperiodic poling period for broadband phase matching. This ensures not only efficient generation of broadband harmonics but also simultaneous phase matching of multiple sum-frequency generation (SFG) processes to ensure that enough optical power is transferred to VIS (third harmonic) and UV (fourth harmonic). 

The waveguide design and dispersive characteristics are similar to those reported in \cite{Ludwig2024}, with minor modifications. Specifically, a 5 mm-long waveguide with 1 um top width incorporates a 4 mm-long periodically poled section with a linearly chirped poling period from 3.9 to 4.5 um. This provides first-order phase matching for second-harmonic generation (SHG) and higher- (third-) order phase matching to third- and fourth-harmonics. 
More details regarding the fabrication of the periodically-poled lithium niobate waveguides have been presented in \citet{Ayhan2025}.

The optical pulses are injected into the waveguide using free-space optics, shown in Figure~\ref{Fig:OpticalChip}, with estimated input coupling losses of \VAL{$6\,\mathrm{dB}$}. 
The input coupling is actively stabilized in X and Y axes to ensure optimal light injection into the waveguide and stable alignment. After the chip, the light is split by a \VAL{20:80} beamsplitter: the reflected \VAL{$20\,\%$} of the light are band-pass filtered around \VAL{$700\,\mathrm{nm}$} and detected with a silicon photodiode, which serves as a signal for the feedback loop adjusting the position of the input beam. The correction signal controls two axes of the 3-axis stage holding the optics. The focus axis, however, is not locked but controlled manually instead.
After the optical chip, the resulting broadband spectrum is collected using an off-axis parabolic mirror and injected into a \VAL{$50\,\mathrm{\mu{}m}$} multimode optical fiber.

\subsection{Mode Scrambler}
\label{Sec:ModeScrambler}

\begin{figure}
 \includegraphics[width=\linewidth]{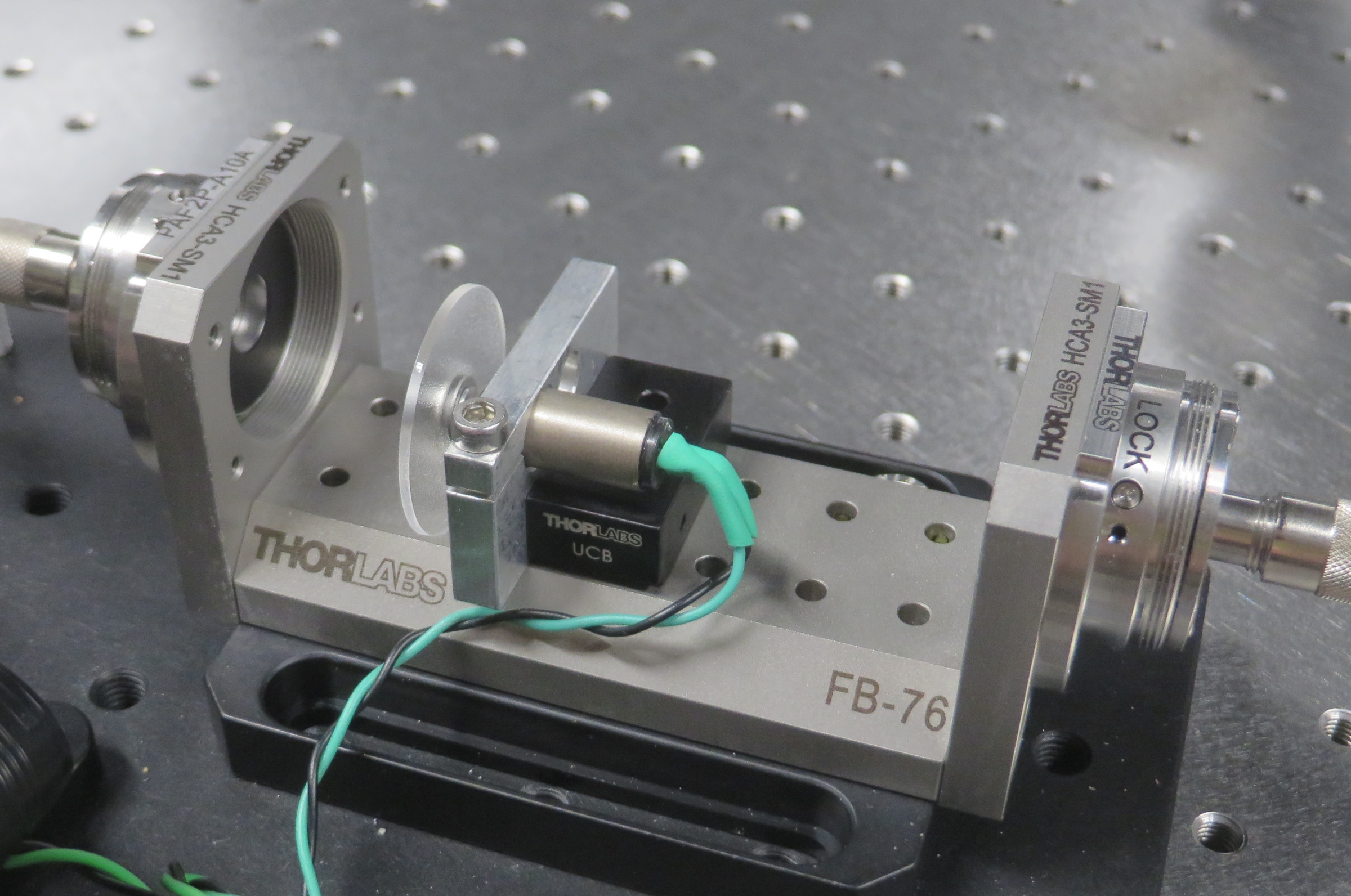}
 \caption{Mode scrambler used to couple the coherent laser light into large multi-mode fibers feeding the calibration unit. The system is composed of two fiber collimators (right and left), with a rotating diffuser disc in between.}
 \label{Fig:ModeScrambler}
\end{figure}

Directly feeding light from a coherent laser source via single-mode or small multimode fibers to a spectrograph designed around large multimode fibers, matched to the atmospheric seeing, will lead to modal noise and severe systematics in the spectrum \citep[e.g.][]{Baudrand2001,Chazelas2010, Chazelas2012, Oliva2019}.
To overcome this issue and produce a homogeneous and\,--\,at least on average\,--\,equal population of all modes in the spectrograph fibers, a dynamic mode-scrambler was used.
The system, shown in Figure~\ref{Fig:ModeScrambler}, is a substantial evolution of the device developed for \citet{Obrzud2018} and almost identical to the one used by \citet{Ludwig2024}. It is composed of a free-space setup in which the input fiber beam, coming from the laser setup, is first collimated, sent through a rotating diffuser disc, and then re-focused into the output fibers. The scattering by the diffuser disc causes an enlarged spot in the focal plane, substantially overfilling the output fiber cores. Movement of the diffuser then breaks the otherwise static relation between the laser speckles and leads to an illumination pattern that very quickly averages-out.
In the output focal plane, a fiber-bundle with two cores of \VAL{$320\,\mathrm{\mu{}m}$} diameter each was placed. These two octagonal core fiber were then connected to the A and B inputs of the ESPRESSO calibration unit. The throughput of the scrambler for each of the two outputs was \VAL{$\approx10\,\%$}.

\subsection{Spectral Flattener}

In addition, a simple spectral flattener setup was constructed to be able to balance the LFC intensity in the various wavebands.
It was designed around a digital light processing (DLP) micromirror array, in which each pixel can be flipped with high frame-rate to either an \textit{on} or \textit{off} position.
Light from the input fiber was dispersed using a low-resolution grating spectrometer, selectively reflected by the DLP mirror, and re-combined into the output fiber by the same spectrometer used in double-pass.
Placed in between the LFC setup and the mode-scrambler, it unfortunately had a throughput of just a \VAL{few per-cent}, significantly lengthening the required exposure times, and covered only wavelengths from the blue end of the ESPRESSO range to about \VAL{$6800\,\AAA$}, rendering the second harmonic of the laser unusable.
The main issue for usage of the flattener, however, was of operational origin. Despite being located directly next to ESPRESSO, no real-time access to the raw frames was possible. Instead, all data first had to be uploaded to the ESO archive in Garching and after they appeared there manually re-downloaded to Paranal. This process, hampered by slow internet and WiFi connections on the mountain, took between 15~minutes and 2.5~hours, far too long for an automated on-the-fly adjustment of the spectral flattener to the flux levels observed by the spectrograph. Manually changing the flattener configuration based just on the visual inspection of the raw frames proved as well too cumbersome and unpractical.
For these reasons, almost all spectra presented here were taken without the flattener, benefiting from the substantially higher throughput and full spectral range, and relying on the comparatively large dynamic range of ESPRESSO.

\section{Basic Spectral Properties}

\begin{figure*}
 \includegraphics[width=\linewidth]{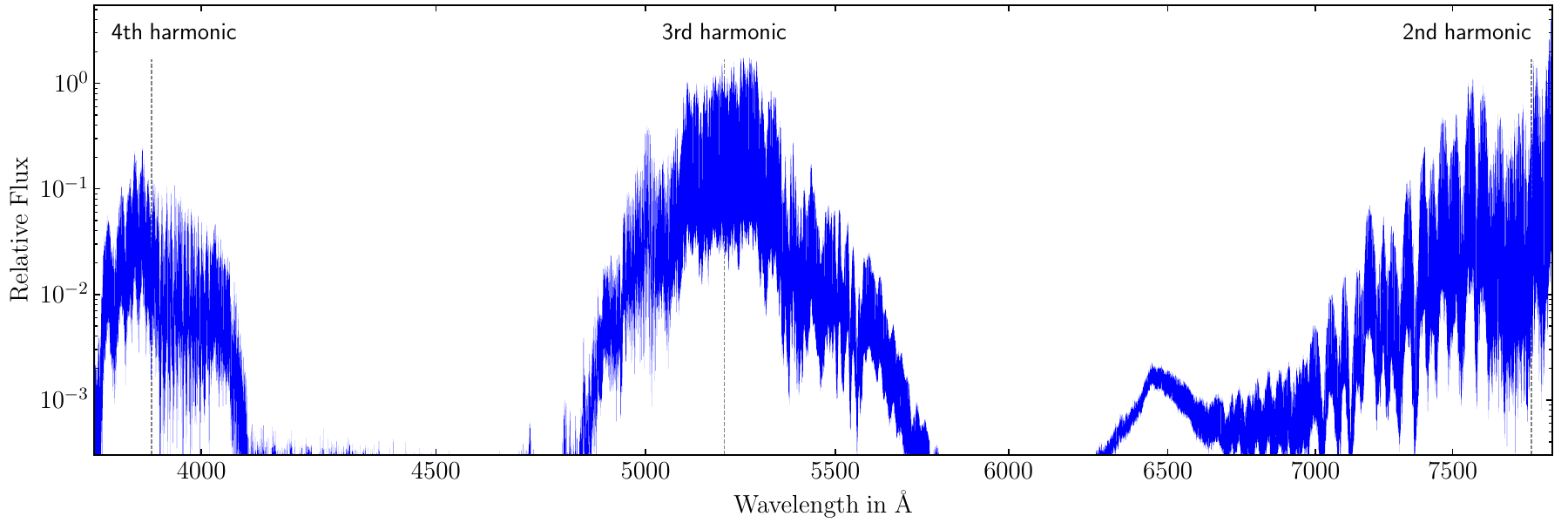}
 \caption{Overview of a spectrum from our $34\,\mathrm{GHz}$ EOM comb, showing the full spectral range covered by ESPRESSO. Flux is given relative to the flatfield calibration source. Vertical dashed lines indicate the centers of the fourth, third, and second harmonic. Magnified views of small portions of the spectral range, clearly showing the comb structure of the spectrum, are presented in Figures~\ref{Fig:LineShape} and \ref{Fig:Sequence_2024-12-09_PhaseMatching}.}
 \label{Fig:SpectrumBLUVES}
\end{figure*}

An example spectrum produced by our $34\,\mathrm{GHz}$ EOM comb is shown in Figure~\ref{Fig:SpectrumBLUVES}. The flux is given in arbitrary unit, relative to the flatfield calibration source. A quantitative flux calibration is hardly possible since the light paths through the telescope, utilized for the observation of spectrophotometric standard stars, and the path via the calibration unit used to feed the LFC to the spectrograph, differ in throughput by a large, unknown factor.
The figure shows the full spectral range of ESPRESSO. At this scale, individual comb lines are not distinguishable, but zoom-ins are presented as part of the following sections, focusing on detailed properties of the LFC spectrum, e.g. in Figure~\ref{Fig:LineShape}. Figure~\ref{Fig:SpectrumBLUVES}, as a general overview, however, clearly depicts the concept of harmonic generation that the LFC setup relies upon. Starting from the seed laser at \VAL{$1560\,\mathrm{nm}$}, the frequencies are doubled, tripled, and quadrupled. These nominal center wavelengths of the harmonics are indicated in Figure~\ref{Fig:SpectrumBLUVES} and one can clearly notice the broadening of the comb around these. For the second harmonic, only the blue half of the harmonic is covered by the ESPRESSO spectral range, while the long-wavelength side falls outside the detector.

The exposure time for the spectrum shown in Figure~\ref{Fig:SpectrumBLUVES} was \VAL{$25\,\mathrm{s}$} and the flux in the third and second harmonic reaches levels comparable to the flatfield source, which corresponds to approximately \VAL{half the full-well capacity} of the detectors.
The spectrum shows clearly detectable flux roughly between \VAL{$3800$} and \VAL{$4080\,\AAA$}, from \VAL{$4880$} to \VAL{$5700\,\AAA$}, and beyond \VAL{$6800\,\AAA$}. The bump in the spectrum around \VAL{$6450\,\AAA$} contains significant flux, but exhibits very poor contrast of the lines and is therefore not usable.
Also, one notices that the spectrum is strongly structured with numerous small regions, predominantly in the wings of the harmonics, where the flux drops significantly. Nevertheless, we identify over \VAL{6000} unique LFC lines.
The total RV information, or RV photon-noise uncertainty, in this spectrum is about \VAL{$37\,\cmps$} per Fiber and Slice in the fourth harmonic and \VAL{$5$} and \VAL{$6\,\cmps$} in third and second harmonic, respectively.
We stress that the spectrum shown in Figure~\ref{Fig:SpectrumBLUVES} was taken without a spectral flattener. It reflect the intrinsic flux levels produced by the comb, combined with the throughput and efficiency of the spectrograph, including calibration unit and fiber link.

In the following sections, we present a more thorough analysis of the spectrum produced by the $34\,\mathrm{GHz}$ EOM comb and analyze in detail several aspects that define its capability to facilitate a precise, accurate, and stable calibration of astronomical high-resolution echelle spectrographs.

\section{Spectrally-Diffuse Background}
\label{Sec:DiffuseBackground}

 A common problem often present in LFC spectra is a spectrally-diffuse background \citep[see. e.g.][]{Blackman2020,Schmidt2021,Schmidt2024}.
 It represents a substantial nuisance for the analysis, can be quite variable in time%
 \footnote{Relative variations of the background fraction in individual spectral orders of up to $\pm$50\,\% have been observed previously in spectra of the ESPRESSO facility LFC. An analysis of the variation of the background fraction in our 34\,GHz EOM comb is presented later in Section~\ref{Sec:Stability}, revealing relative variations up to $\pm$30\,\%.}%
 , has thus to be determined in each LFC spectrum, modeled, and removed before measuring the lines or alternatively be fitted simultaneously.
 Excessive background flux might also dominate the overall photon noise in the LFC spectra and lead to a reduction of the achievable precision. To avoid this, the flux in the actual LFC lines should contain substantially more photons than the spectrally-diffuse background.
 As a quantitative measure for this, we define the background fraction as
 \begin{equation}
  f_\mathrm{BG} = \frac{ \int_\mathrm{FSR} F_\mathrm{BG}(\lambda) \: \mathrm{d}\lambda }{ \int_\mathrm{FSR} F_\mathrm{line}(\lambda) \: \mathrm{d}\lambda }\;.
  \label{Eq:BackgroundFraction}
 \end{equation}
 Here, $F_\mathrm{line}(\lambda)$ and $F_\mathrm{BG}(\lambda)$ denote the flux in the LFC lines and in the spectrally-diffuse background. To compute these in practice, one requires separate models of the background and line fluxes, obtained e.g. by interpolating the minima in between lines and fitting appropriate profiles to the lines. The integrals shall run over one (or multiple) free-spectral range (FSR).
 We stress that in this definition, $f_\mathrm{BG}$ is independent of the spectrograph resolution.
 According to the goal stated above, it should be $f_\mathrm{BG} \ll 1$.

 Our test of a similar $18\,\mathrm{GHz}$ EOM comb in 2023 at OHP/SOPHIE \citep{Ludwig2024} revealed a strong dependence of the spectrally-diffuse background flux on wavelength. In accordance with the concept of harmonic-generation, the background fraction is lowest in the centers of the harmonics and then grows quadratically with separation from the harmonic center. This could be empirically inferred from the data, but also explained by a model based on the propagation of the phase-noise in the laser pulses, originating from the RF oscillator \citep[see supplementary material of ][]{Ludwig2024}.
 The goal here is to improve over the OHP/SOPHIE results and demonstrate the suppression of the spectrally-diffuse LFC flux using an adequate noise filtering strategy, in particular by a high-finesse FP filter cavity introduced in the laser setup \citep[see e.g.][]{Kashiwagi2016, Beha2017}.
 For this test with ESPRESSO, we are also able to measure the spectrally-diffuse background in the LFC spectra much better and with substantially less complications than for the OHP/SOPHIE data. Beneficial is the larger separation of the LFC lines ($34\,\mathrm{GHz}$ instead of $18\,\mathrm{GHz}$) and the much higher resolving power of ESPRESSO ($R=\frac{\lambda}{\Delta{}\lambda}\approx135\,000$ compared to $75\,000$ for SOPHIE). Determination of the background level in the LFC spectra is therefore readily and unambiguously possible over the full wavelength range of ESPRESSO, also in the near-UV.

 \begin{figure}
  \includegraphics[width=\linewidth]{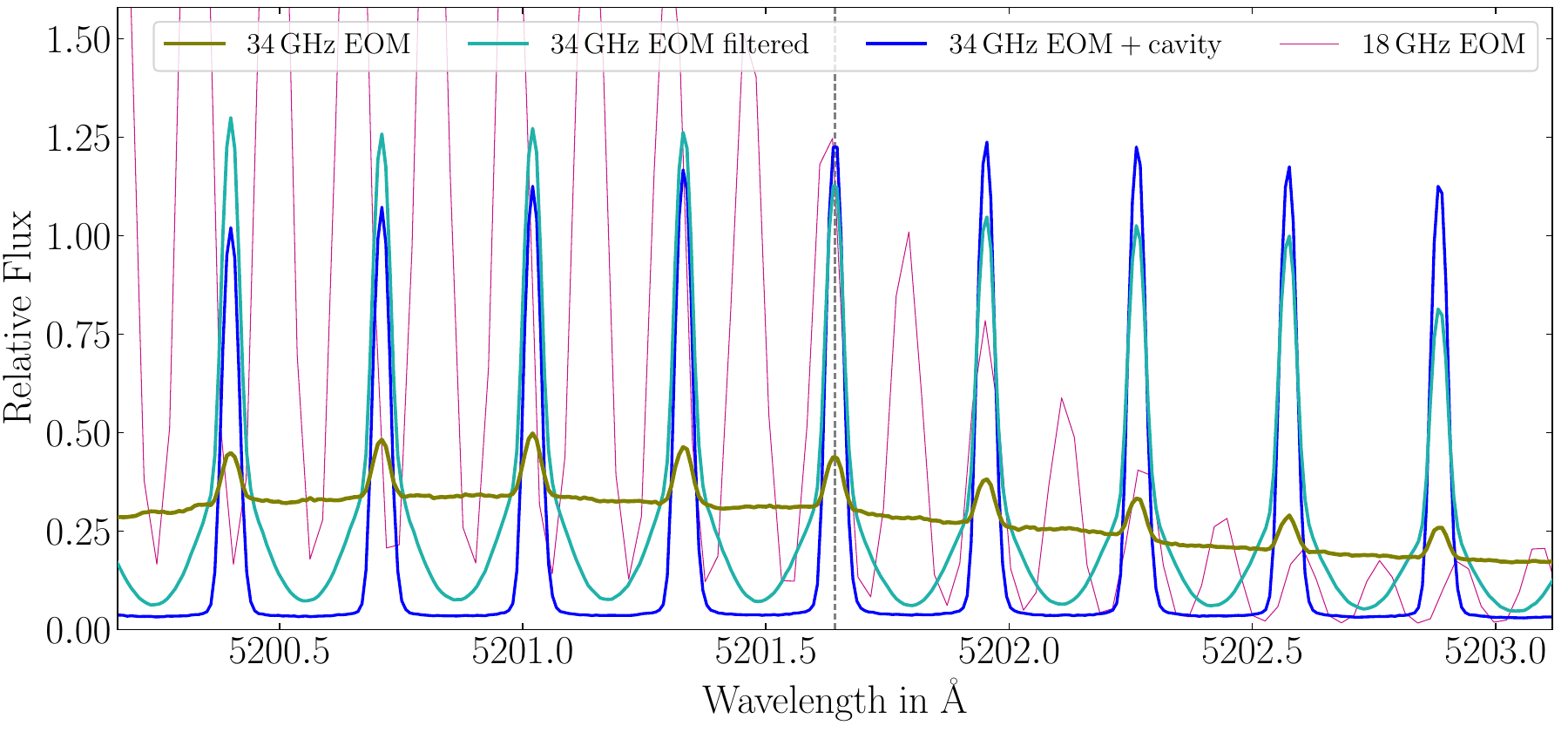}%
  \caption{
  ESPRESSO spectra of the 34\:GHz EOM comb obtained with different strategies for noise filtering: no filtering at all (\VAL{olive}), a tuned waveshaper mask (\VAL{cyan}), and a locked high-finesse FP cavity (\VAL{blue}).
  A small section around the center of the third harmonic (vertical dashed line) is shown.
  For comparison, the spectrum from our 2023 test of a 18\,GHz EOM comb at OHP/SOPHIE \citep{Ludwig2024} is shown in \VAL{magenta}. Due to the different setup, neither the lines nor the center of the harmonic do line up.
  }
  \label{Fig:LineShape}
 \end{figure}

 We tested the LFC in three different configurations: without any specific means for noise filtering, using a waveshaper mask, and with the FP cavity.
 The results of this are visualized in Figure~\ref{Fig:LineShape}, which shows a small spectral section around the center of the third harmonic.
 For the first spectrum (shown in \VAL{olive}), no filtering was employed at all. Obviously, the spectrum is dominated by spectrally-diffuse background flux with just a small fraction of the total photons being located in the actual lines.

 \begin{figure}
  \includegraphics[width=\linewidth]{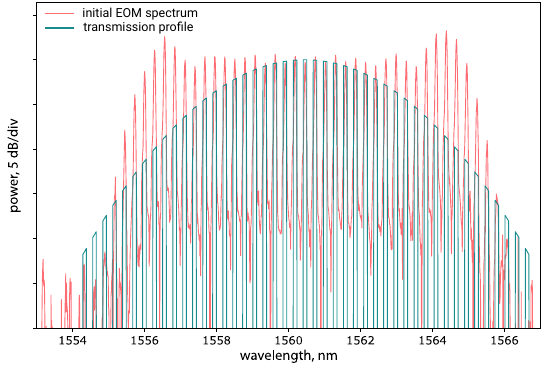}%
  \caption{
   Visualization of the fundamental comb spectrum and the applied waveshaper mask used to suppress noise and background in-between the comb modes.
  }
  \label{Fig:WaveshaperMask}
 \end{figure}

 For the second spectrum, shown in \VAL{cyan}, a mask on the waveshaper was applied that suppresses a significant part of the flux in between the lines of the fundamental EOM comb. This mask, together with the spectrum of the fundamental comb, is visualized in Figure~\ref{Fig:WaveshaperMask}. The mask consists of a regular pulse shaping, defined by the envelope of the mask, required for Gaussian pulse generation. In addition, each individual comb mode is filtered to suppress the noise and background in between the individual modes.
 The spectral resolution of the waveshaper, however, is too coarse to facilitate a very high degree of noise filtering, but nevertheless is able to suppress a certain fraction of it.
 With this setup, the output spectrum exhibits after broadening a rather peculiar shape: the relatively narrow lines sit on top of a pyramidal base.
 Nevertheless, the applied waveshaper mask substantially reduces the total amount of background flux and  leads to a strong increase in the number of photons that end up in actual comb lines after the broadening and harmonic-generation process.
 The unfiltered spectrum and the one employing a waveshaper mask were otherwise taken with identical parameters.

 The final spectrum (displayed in \VAL{blue} in Figure~\ref{Fig:LineShape}) is the one with the nominal setup, utilizing a high-finesse FP cavity ($\VAL{\mathcal{F} \approx 725}$) for noise filtering before the final amplification and spectral broadening stages. Clearly, this drastically reduces the spectrally-diffuse background and the ESPRESSO spectrum displays sharp and narrow lines with a flat and low background in between them.

 \begin{figure*}
 \includegraphics[width=\linewidth]{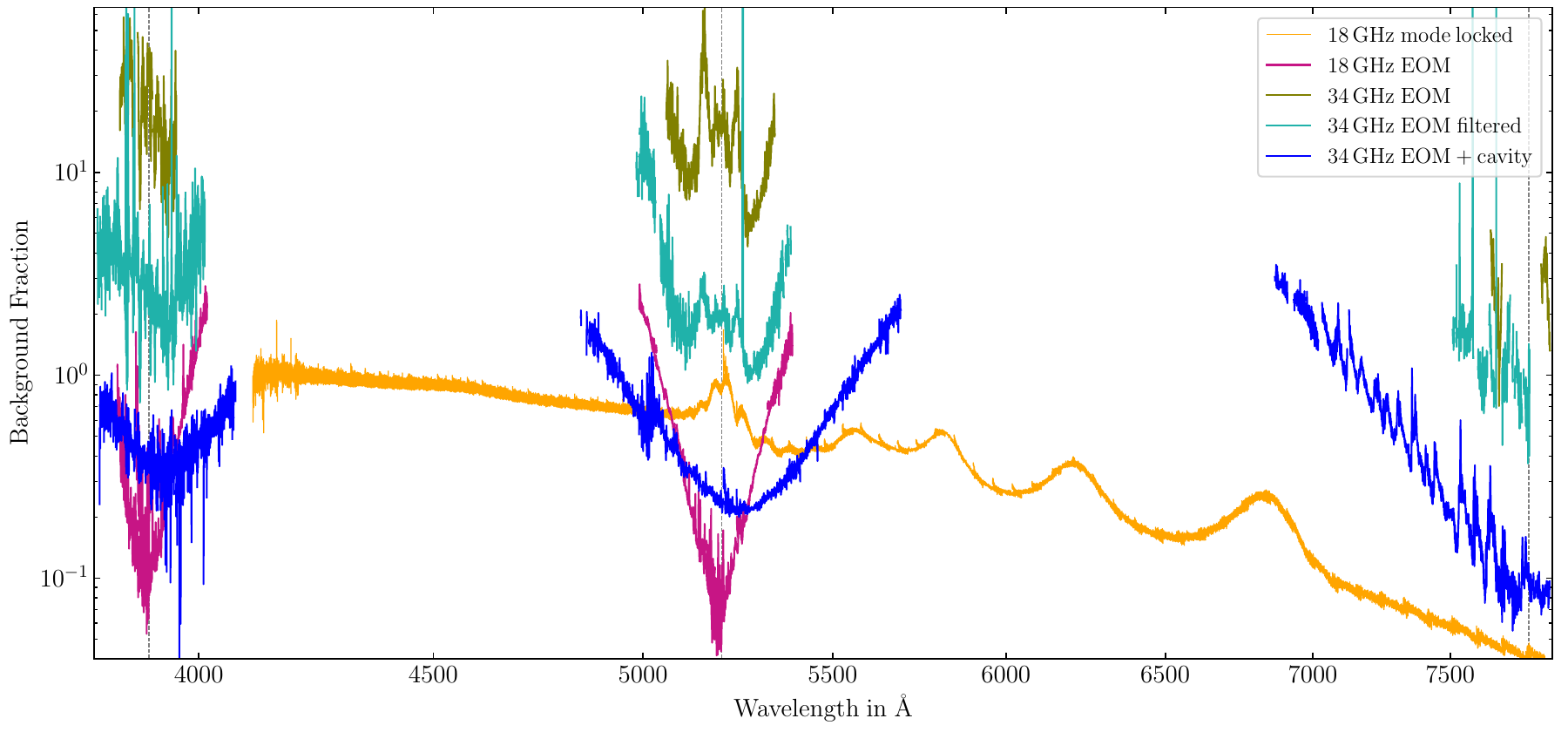}%
 \caption{
  Background fraction, defined as the ratio between the flux contained in the spectrally-diffuse background and in the LFC lines, for various comb configurations, i.e.
  the 34\:GHz EOM comb without any filtering (\VAL{olive}), using a tuned waveshaper mask (\VAL{cyan}), and a locked high-finesse FP cavity (\VAL{blue}) for noise suppression.
  Also shown is the background fraction from our 2023 test of a $18\,\mathrm{GHz}$ EOM comb at OHP/SOPHIE \citep[\VAL{magenta},][]{Ludwig2024}.
  The vertical dashed lines indicate the centers of the harmonics.
  In addition, the same quantity is also shown for the ESPRESSO $18\,\mathrm{GHz}$ mode-locked laser, which is defined around $1040\,\mathrm{nm}$ and directly broadened all the way to the visible regime.
  }
 \label{Fig:BackgroundFraction}
\end{figure*}

 In Figure~\ref{Fig:BackgroundFraction}, we quantitatively compare the amount of background flux in the different spectra and assess the dependence with wavelength. The plot shows the background fraction, as defined in Equation~\ref{Eq:BackgroundFraction}, for the different setups of the 34\:GHz EOM comb, the 18\:GHz EOM comb tested at OHP/SOPHIE \citep{Ludwig2024}, and the ESPRESSO 18\:GHz mode-locked facility LFC. For the spectrum created by filtering with a waveshaper mask, we only consider the narrow core as line flux, while we consider the pedestal as well as an additional flat component as background. The reasoning behind this is that only the line itself is locked in frequency while the pedestal is a product of the noise processes and probably subject to wavelength drifts. It can therefore not be used to derive reliable wavelength information.

 The EOM combs using harmonic-generation in TFLN waveguides show a quite characteristic pattern: The background fraction is lowest in the centers of the harmonics and increases quickly with increasing broadening away from the center.
 This is most visible for the 18\:GHz EOM comb. As given in \citet{Ludwig2024},  the amount of background should increase approximately quadratically with mode number. Given the logarithmic scaling in Figure~\ref{Fig:BackgroundFraction}, this appears as sharp V-shape.

 For the unfiltered 34\;GHz EOM comb, the amount of background is\,--\,as already shown in Figure~\ref{Fig:LineShape}\,--\,much higher.
 In addition, it does not follow exactly the same pattern. As expected, far away from the harmonic, the background fraction increases rapidly and also with a similar slope as for the 18\;GHz EOM comb (best visible for the third harmonic, where the data quality is the best). In the center, however, it does not reach a minimum. Instead, there appears to be an additional component, composed of a central and two side peaks.
 The side peaks are separated by \VAL{$\approx80\,\AAA$}, corresponding to either \VAL{$8.8\,
 \mathrm{THz}$} or approx. \VAL{260} comb modes. 

 Filtering by the waveshaper substantially improves the background, but this effect has no strong wavelength dependence. Instead, the background fraction is globally reduced by a roughly constant factor.
 The improved line flux, however,  allows to measure the background over a wider spectral range.

 Suppressing the noise with the high-finesse FP cavity allows to further improve the background.
 As shown in Figure~\ref{Fig:BackgroundFraction}, roughly two orders of magnitude can be gained compared to the unfiltered case.
 One also notices that the background fraction increases with a shallower slope away from the center of the harmonic and thus enables a much wider spectral range to be used. Still, the observed background fraction in the center of the harmonics is nevertheless higher than for our $18\,\mathrm{GHz}$ EOM comb tested at OHP/SOPHIE \citep{Ludwig2024} and appears to increase from second, to third, to fourth harmonic.

 If one applies a requirement of $f_\mathrm{BG} \leq 1$, the $34\,\mathrm{GHz}$ EOM comb equipped with the filtering cavity provides in the fourth harmonic coverage from \VAL{$3900$} to \VAL{$4070\,\AAA$}, actually limited by the flux in the shown exposure. This corresponds to \VAL{$-2.6\,\%$} and \VAL{$+4.3\,\%$} relative bandwidth around the center of the harmonic. For the third harmonic, we find a background fraction smaller than unity within the spectral range from \VAL{$4930$} to \VAL{$5560\,\AAA$}, corresponding to relative bandwidths of \VAL{$-5.2\,\%$} and \VAL{$+6.9\,\%$}. The second harmonic is only partially covered by the ESPRESSO spectral range, allowing to only state a lower limit around \VAL{$7160\,\AAA$}, corresponding to a relative bandwidth of \VAL{$-8.2\,\%$}.


 Figure~\ref{Fig:BackgroundFraction} shows in addition also the background fraction of the $18\,\mathrm{GHz}$ mode-locked facility LFC. Due to the vastly different design concept for the spectral broadening, the shape is very different. This comb is created at a central wavelength close to 1040\,nm \citep{Probst2016} and then, instead of harmonic generation in PPLN waveguides, directly broadened in a tapered photonic crystal fiber all the way into the visible regime. Accordingly, the background fraction is the lowest at the long-wavelength end of the ESPRESSO spectral range and increases basically monotonically towards shorter wavelength.
 However, there are additional features, most importantly some humps between \VAL{5500} and \VAL{$7000\,\AAA$}, which are not stable in time but might change from exposure to exposure, as well as some feature around \VAL{$5200\,\AAA$} that probably results from the seed laser.

 The data presented here clearly shows the dramatic improvement in the spectrally-diffuse background that can be achieved by introducing filtering cavities for noise suppression. However, these devices also come with additional operational complexity. In particular, the FP cavity has to be locked to the seed laser.
 Locally, over a small spectral range, the modes of the FP cavity follow a simple, affine relation, equivalent to the comb equation presented before in Equation~\ref{Eq:Comb}.
 To achieve a good match, FSR and offset should over the spectral range of the fundamental comb, in our case roughly \VAL{$\pm20$} modes, be identical for the cavity and the LFC. However, these two parameters are typically not independently controllable for the cavity.
 The locking scheme basically ensures that the central comb mode lines up with one of the FP transmission peaks.
 However,  despite the locking to the seed laser, we noticed that over the duration of the experiment the FSR of the cavity varied noticeably with temperature.
 This can\,--\,in the simplified picture outlined above\,--\,be explained assuming that the effective offset frequency of the cavity, or equivalently the phase shift introduced by it, varies with temperature and the locking scheme then adjusts the FSR to again line up the central mode of the comb with a cavity mode.
 In such a situation, the only practical solution to achieve a good transmission of the full fundamental comb is to re-adjust the repetition rate of the LFC to match the FSR of the cavity, and we did this manually.
 The relative changes of the repetition rate were small, of the order of \VAL{$10^{-4}$}, and, as long as repetition rate and offset frequency of the comb are accurately known, this poses no real limitation for the wavelength calibration, one just has to properly take into account the actual repetition rate for each LFC exposure.
 However, it means that one no longer has the freedom to choose both comb-defining frequencies at will or keep them exactly constant. Instead, one is subject to environmental effects and has to adapt the comb accordingly. In consequence, this also mandates a constant forwarding of the actual repetition rate and offset frequency from the LFC setup to the data reduction system.
 These aspects have also been described in detail by \citet{Sekhar2024} within the context of LFC tunability for a rather similar EOM comb setup.

\section{LFC Line Width}
\label{Sec:LineWidth}

\begin{figure*}
 \includegraphics[width=\linewidth]{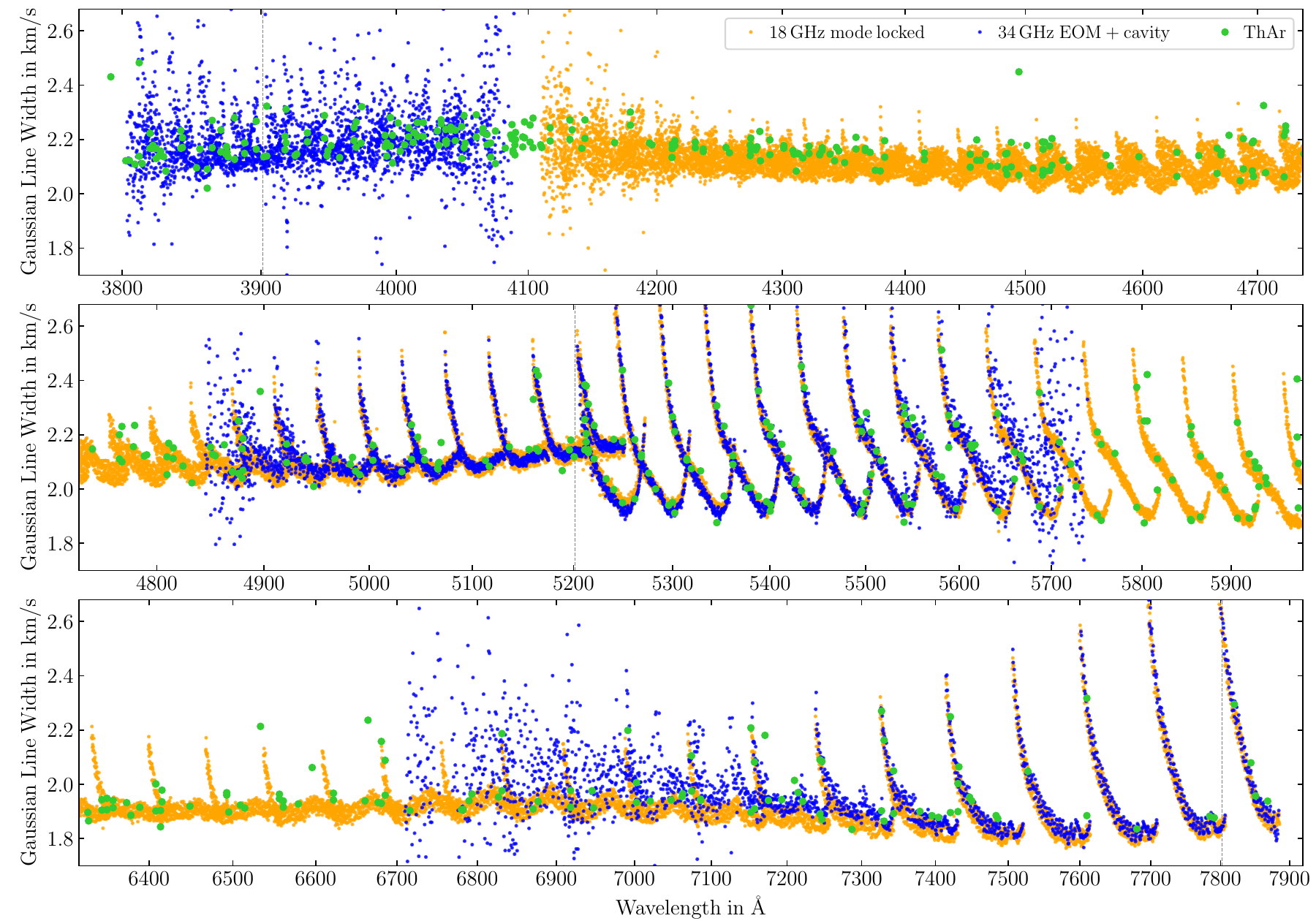}%
 \caption{Apparent line width of different ESPRESSO calibration sources as function of wavelength. Given is the FWHM obtained by fitting Gaussian profiles to the individual lines. The quasi-periodic modulation reflects the variation of the spectral resolution along the spectral orders and is typical for echelle spectrographs \citep[see e.g.][]{Schmidt2024}. Apart from this, subtle differences in the line with can be noticed for the different sources. The centers of fourth, third, and second harmonic are indicated by vertical dashed lines. The transition between blue and red spectrograph arms are clearly visible between 5230 and 5270 $\mathrm{\AAA}$.
 }
 \label{Fig:LineWidth_Harmonics}
\end{figure*}

Our model for the propagation of the phase noise presented in \citet{Ludwig2024} did not only predict a raising background fraction with increasing spectral broadening, but also a growing line width. This is of course undesired. The LFC shall deliver narrow and unresolved lines to facilitate an accurate and stable wavelength calibration. In addition, negligible line width is also a strict requirement when using the LFC for characterizing the instrumental LSF.
However, during our previous measurement campaign at OHP/SOPHIE it was, due to the limited spectral resolution of the instrument, not possible to measure the intrinsic LFC line width.
For this test, ESPRESSO offers in this regard much better capabilities. First, its resolving power is about twice as high. More importantly, the availability of the facility LFC allows an a~priori characterization of the LSF, which makes it possible to determine intrinsic line widths far narrower than the approximately $2.2\,\kmps$ wide instrumental LSF \citep[see][]{Schmidt2024}. The achievable precision is of course not comparable to heterodyne beat-note measurements, but simultaneous measurements of all lines in the spectrum are possible.

In the following, we avoid the complications of modeling the LSF and instead directly compare the apparent line width of various sources. This is substantially more robust than inferring the almost vanishing intrinsic width of unresolved lines.
Figure~\ref{Fig:LineWidth_Harmonics} thus shows the apparent FWHM of all lines in the ThAr spectrum, the $18\,\mathrm{GHz}$ mode-locked laser, and our $34\,\mathrm{GHz}$ EOM comb configured with the high-finesse FP cavity for noise suppression. To avoid confusion, only data extracted from Slice~a of Fiber~A is shown. The apparent line widths are obtained by fitting a Gaussian profile, in addition to first-order polynomials, to each line in question. A Gaussian profile is clearly no accurate description of the line profile, but it is still allows to determine a reasonable measure for the line width.

The most apparent feature in Figure~\ref{Fig:LineWidth_Harmonics} is the semi-periodic modulation that relates to the echellogram structure of ESPRESSO. It reflects the change of the instrumental resolution along individual echelle orders \citep{Schmidt2024}, because the intrinsic line width of all three sources is far below the resolution of the spectrograph.
The lines of the $18\,\mathrm{GHz}$ mode-locked laser are supposed to be extremely narrow \citep[$\approx120\,\mathrm{kHz}$,][]{Probst2014} and should therefore directly reflect the instrumental LSF.
The thorium lines are in general slightly wider than the lines of the $18\,\mathrm{GHz}$ mode-locked LFC, in accordance to the findings in \citet{Schmidt2024} and their origin from a HCL. However, ThAr lines are not necessarily a clean measurement due to varying intensity, blends, and even saturated lines.

Line widths of the $34\,\mathrm{GHz}$ EOM comb are in the central parts of the harmonics fully consistent with the ones from the $18\,\mathrm{GHz}$ mode-locked laser and therefore have to be considered fully unresolved. For the fourth harmonic, in the near-UV, no comparison to the other LFC is possible. Nevertheless, also in this region, the LFC lines appear of equal width or narrower than the thorium lines, demonstrating that they are also here unresolved.
However, with increasing distance from the center of the harmonics and thus more spectral broadening, the LFC lines start to exhibit a measurable width. This can be seen e.g. in the second harmonic for \VAL{$\lesssim 7400\,\mathrm{\AAA}$}, where they start to become clearly wider than the lines of the mode-locked LFC (Figure~\ref{Fig:LineWidth_Harmonics}, bottom panel). Similar effects can also be seen in the third harmonic for \VAL{$\lesssim 4950\,\mathrm{\AAA}$} and \VAL{$\gtrsim 5600\,\mathrm{\AAA}$} (Figure~\ref{Fig:LineWidth_Harmonics}, central panel). This is in full (qualitative) agreement with the phase-noise based model of the line width presented in \citet{Ludwig2024}.

\begin{figure}
 \includegraphics[width=\linewidth]{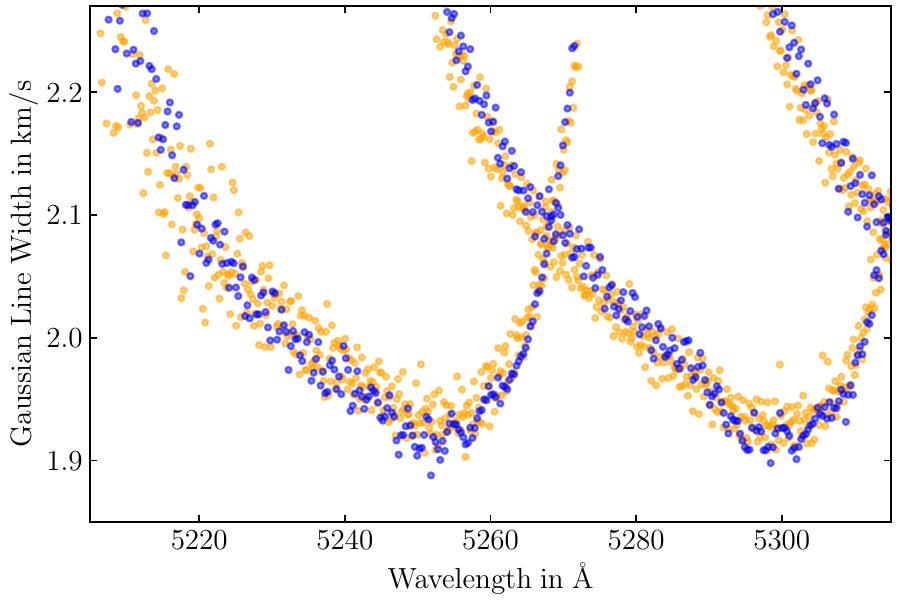}%
 \caption{Magnified view of Figure~\ref{Fig:LineWidth_Harmonics}.}
 \label{Fig:LineWidth_HarmonicsZoom}
\end{figure}

A very close inspection of the observed line widths for the third harmonics (Figure~\ref{Fig:LineWidth_Harmonics}, central panel) suggests that in the center of the harmonic, e.g. in spectral order 116 around \VAL{$5250\,\mathrm{\AAA}$}, the lines of the EOM comb might actually appear slightly narrower than the ones of the mode-locked laser.
A zoom-in of this region is presented in Figure~\ref{Fig:LineWidth_HarmonicsZoom}.
The difference is tiny and less than \VAL{1\%} of the line width.
It can therefore not be fully excluded that this apparent difference might be related to some systematics in the measurement process. However, if actually true, it would be consistent with the expected behavior.
The mode-locked laser is based in the near-IR at $\approx 1040\,\mathrm{nm}$ and there, before the spectral broadening, exhibits a vanishing line width of just $120\,\mathrm{kHz}$, as demonstrated in \citet{Probst2014}.
Nevertheless, the tremendous spectral broadening into the visible spectral range might indeed lead to an increase of the line width, following the model presented in \citet{Ludwig2024}.
The EOM comb contains in general substantially more phase noise, even after filtering by the FP cavity, which leads to the quite noticeable increase in line width away from the centers of the harmonics. Despite this, the impact of phase noise should, due to the adopted concept of harmonic-generation in the EOM comb setup, be very low and the lines narrow at the center of each harmonic.
It might therefore very well be that we observe this effect\,--\,an increase of line width\,--\,in the spectra of both LFCs, however, with very different spectral dependencies.
The observations thus, at least qualitatively, match our understanding and expected behavior, but the magnitude of the effect is so minute that detecting it pushes the limits of grating spectrographs, even for ESPRESSO. A clear and definitive answer would thus require heterodyne line width measurements in the optical regime.


\section{LSF Determination}
\label{Sec:LSFDetermination}

\begin{figure*}
 \includegraphics[width=\linewidth]{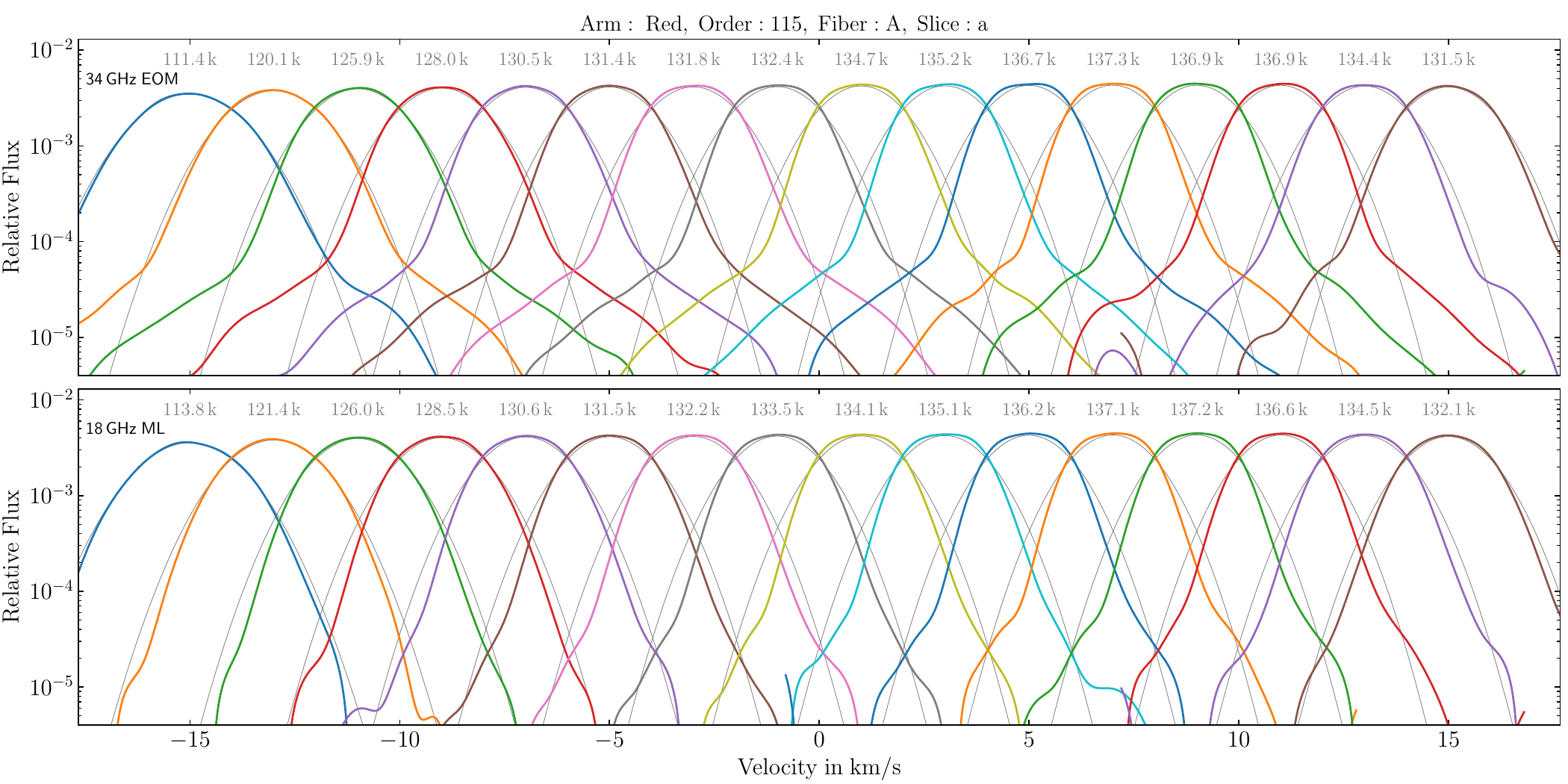}
 \caption{Different ESPRESSO instrumental LSF profiles of one single echelle order (Order\:115, Fiber\:A, Slice\:a) inferred from the 34\:GHz EOM comb (top) and the 18\:GHz model-locked comb (bottom). The different LSF profiles are offset horizontally for clarity and the flux axis is scaled logarithmic to better visualize the extended, low-level wings of the LSF. For comparison, a Gaussian with the same FWHM as the empirical profile is shown in gray. At the top, the resolving power corresponding to that FWHM is labeled. }
 \label{Fig:LSF_LFC-BLUVES}
\end{figure*}

To properly characterize the instrumental LSF, one requires a calibration source that provides a large number of unresolved lines which by themselves contribute negligible intrinsic width and thus reveal the actual instrumental profile. For this reason, the spectra of the $18\,\mathrm{GHz}$ comb were used in \citet{Schmidt2024} to derive an accurate model of the ESPRESSO LSF.
If one tries to define the ideal properties of a light source dedicated to characterization of the instrumental LSF, one notices a conflict between competing design goals:  On the one hand, one would like to have as many lines available as possible, simply to maximize the information content in the spectrum. This demands a close spacing of the lines. However, the separation between lines also limits the extent over which one can determine the LSF. If for instance one wants to characterize extended wings of the LSF, a fairly large spacing is required. In addition, one requires sufficient separation between the lines to accurately and unambiguously measure the spectrally-diffuse background and remove it from the LFC spectrum before inferring the line profile.

Astronomical echelle spectrographs have an approximately constant resolving power $R=\frac{\lambda}{\Delta{}\lambda}$ over the wavelength range, which corresponds to constant resolution in velocity space. For frequency combs, however, lines have a regular spacing in frequency. Thus, the separation of lines in terms of resolution elements varies significantly across the spectral range.
The ESPRESSO $18\,\mathrm{GHz}$ facility LFC provides plenty of separation between lines at the red end of the spectral range, e.g. $14\,\kmps$ or $\gtrsim6\times\mathrm{FWHM}$ at $7800\,\AAA$. At shorter wavelengths of $5000\,\AAA$, however, $18\,\mathrm{GHz}$ corresponds to $9\,\kmps$ and thus only $\approx4\times\mathrm{FWHM}$, clearly showing that the extent over which the LSF can be determined is much smaller.
With the $34\,\mathrm{GHz}$ comb, we are able to characterize the ESPRESSO LSF over nearly twice the width, determine whether there are significant extended wings, and check if modeling these leads to more accurate line fits.

For this, we follow the approach lined out in \citet{Schmidt2024}, based on a Gaussian-process deconvolution algorithm. However, to allow a better characterization of the LSF wings even with fewer available LFC lines, we choose a Gaussian-process kernel that exhibits a variable correlation length, making the prior more stiff in the wings of the LSF but retaining the full flexibility in the core.
In Figure~\ref{Fig:LSF_LFC-BLUVES}, we show for one selected echelle order two sequences of LSF models. They correspond to 16 blocks along the spectral order 115 of ESPRESSO and both sets of profiles are determined in an identical way, however, once extracted from a spectrum of the $18\,\mathrm{GHz}$ mode-locked laser and once from the $34\,\mathrm{GHz}$ EOM comb. Overall, both sets agree well and show\,--\,apart from the unavoidable correlated noise in the models\,--\,the same profile in the core of the LSF. Considering the uncertainties, also the determined FWHMs of the profiles, stated in terms of resolving power above the profiles in Figure~\ref{Fig:LSF_LFC-BLUVES}, are nearly identical. This shows that in general accurate LSF models can be derived from the spectra of both combs.

However, further away than approx. \VAL{$2.5\,\mathrm{GHz}$} from the center of the profile, we find evidence for significant extended wings of the ESPRESSO LSF that are only detectable in the spectrum of the $34\,\mathrm{GHz}$ comb but not in the $18\,\mathrm{GHz}$ one. Here, the LSF profile starts to decay much slower than in the core. Nevertheless, this extended component of the LSF only appears at flux levels far below \VAL{1\,\%} of the peak. It is therefore not immediately clear how relevant this is for modeling ESPRESSO spectra.

\begin{figure}
 \includegraphics[width=\linewidth]{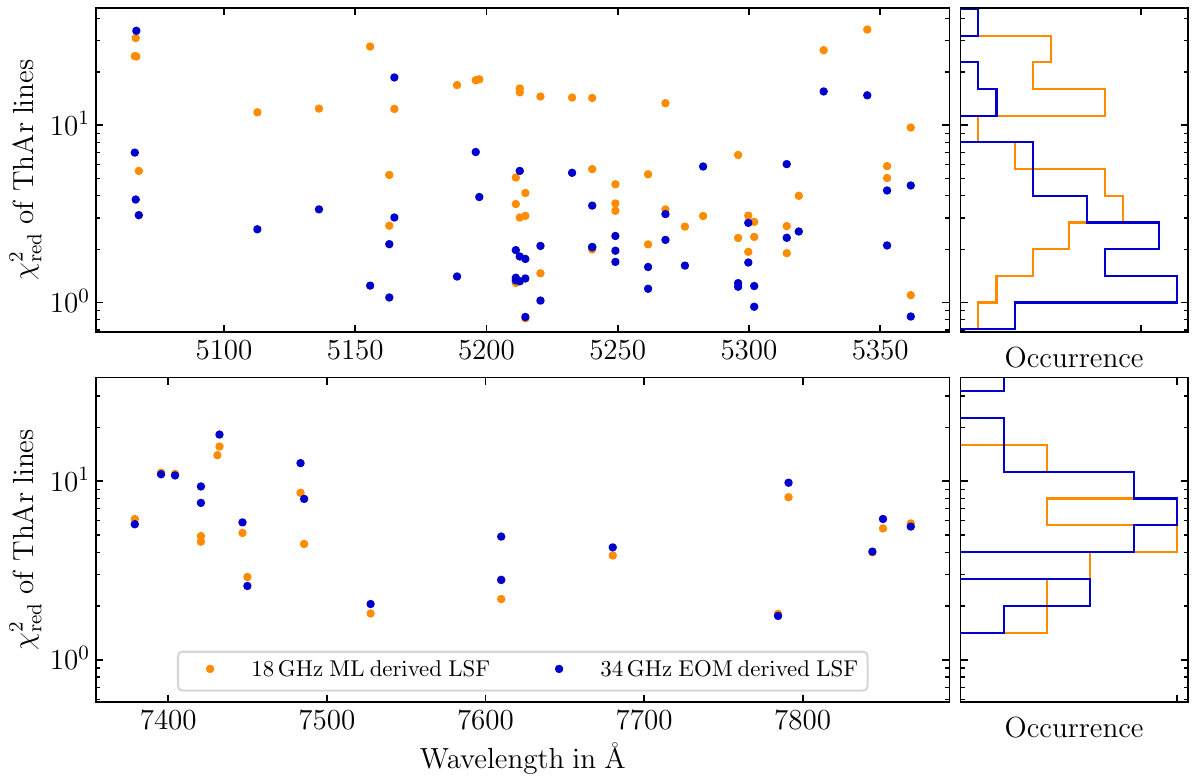}
 \caption{Goodness-of-fit metric for modeling individual lines in the ThAr spectra.
  Different colors indicate the obtained $\chi^2_\mathrm{red}$ for either assuming the LSF inferred from the 18\:GHz model-locked laser or the  34\:GHz EOM comb. The ThAr spectra themselves are in both cases identical. The top panel represent the wavelength range corresponding to the third harmonic, the bottom one the second harmonic. The right panels presents histograms of the $\chi^2_\mathrm{red}$ values.
  }
 \label{Fig:LSF_chi2}
\end{figure}

To assess this aspect and determine which of the two sets of LSF models is more accurate, we use them to fit the thorium lines in the ThAr spectrum. These lines are sparse and relatively isolated. It is therefore possible to fit them over a large spectral window that is sensitive to possible extended wings.
From this test, Figure~\ref{Fig:LSF_chi2} shows for all thorium lines that fall in the spectral range covered by both LFCs the reduced $\chi^2_\mathrm{red}$~metric, representative for the goodness of the fits. For the third harmonic, centered around $5200\,\AAA$, the fits using the LSF model derived from the $34\,\mathrm{GHz}$ comb clearly exhibit a better $\chi^2_\mathrm{red}$, demonstrating that these LSF models are indeed a more accurate description of the ESPRESSO instrumental profile and the extended wings are a real feature and not an artifact of the modeling procedure.
For the second harmonic, at substantially longer wavelengths, both sets of LSF models provide nearly identical $\chi^2_\mathrm{red}$, indicating that here also the $18\,\mathrm{GHz}$ comb provides sufficient line sampling to characterize the full profile.

Despite the better $\chi^2_\mathrm{red}$ statistics, we have to stress that we found no strong evidence that using the LSF models derived from the $34\,\mathrm{GHz}$ comb leads to substantially better wavelength information.
In most cases, e.g. when fitting the FP lines or the $18\,\mathrm{GHz}$ comb, the adopted fitting window anyway has a width not exceeding $\pm4\,\kmps$, over which both sets of model are very similar.

\section{Wavelength Calibration Accuracy}
\label{Sec:Accuracy}

\begin{figure*}
 \includegraphics[width=\linewidth]{ 	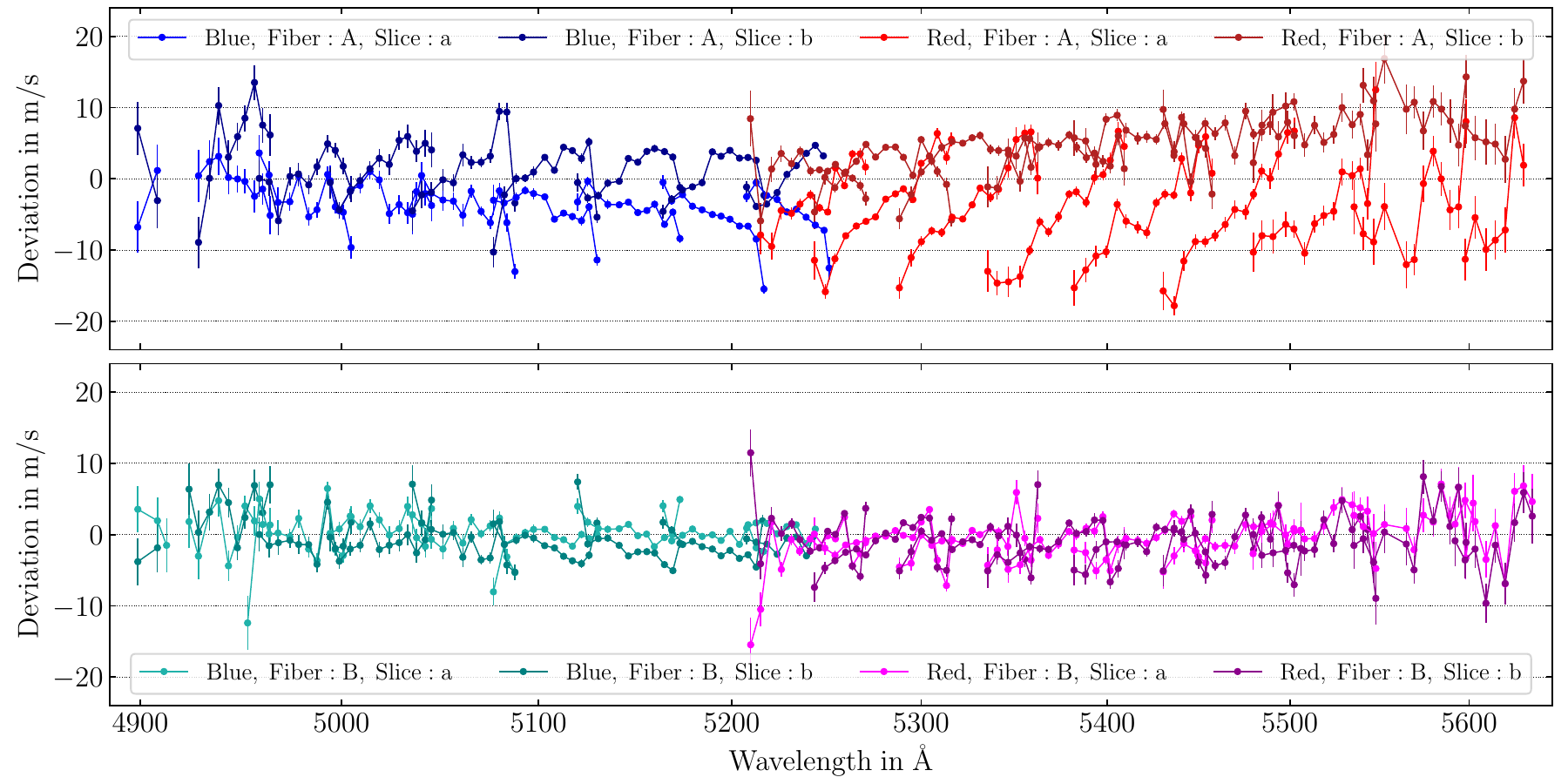}
 \caption{Wavelength calibration accuracy inferred from a comparison of the $34\,\mathrm{GHz}$ EOM comb to the $18\,\mathrm{GHz}$ mode-locked laser. This was obtained by calibrating the instrument with the $18\,\mathrm{GHz}$ comb (LSF and wavelength solution) and then using this to measure the wavelength of the $34\,\mathrm{GHz}$ EOM comb lines. Shown is the difference between inferred and true wavelengths, expressed in velocity units and binned in blocks extending over 1/16th of a spectral order. Colors indicate measurements obtained from blue and red arms of ESPRESSO, the two Fibers, and two Slices. The top panel displays results from Fiber~A, the bottom panel the same for Fiber~B. Shown is the spectral range covering the third harmonic.
 }
 \label{Fig:Accuracy_LFC-BLUVES}
\end{figure*}

 To assess the wavelength calibration \textit{accuracy}, we carry out a careful comparison of both combs. For this, we use the spectra of one comb, in this case the $18\,\mathrm{GHz}$ mode-locked laser, to calibrate the instrument and then, based on this calibration, measure the wavelength of the lines in the other comb spectrum. These can then be compared to their true values defined by the comb equation (Eq.~\ref{Eq:Comb}).
 The result of this is shown in Figure~\ref{Fig:Accuracy_LFC-BLUVES}. To reduce the photon noise, measurements from individual LFC lines are binned in 16 blocks per spectral order, the same as used for determination of the LSF.

 ESPRESSO is equipped with two independent spectrograph fibers, allowing to collect light from a science target and sky simultaneously, or from two calibration sources. In the calibration system, the light from each source is first split in an A and B side and these then channeled via fibers to two independent port-selector units, which allow to chose any of the six input ports and send the light to the spectrograph front-end, where it is injected into the spectrograph fibers (see \citealt{Megevand2014} for details, in particular their Figure~4).
 We have obtained independent wavelength measurements for both spectrograph fibers. In addition, ESPRESSO is equipped with an anamorphic pupil slicer (APSU, \citealt{dellAgostino2014}) that, after the light has exited the fiber train, slices the far-field and independently projects both halfs onto the echelle grating,  producing two traces per input fiber that are extracted and processed independently. Therefore, four wavelength measurements of any line are available.

 Naturally, and independent of the actual light sources, all four of these measurements should be at least consistent with each other and provide the same value. However, significant discrepancies between fibers and slices had already been reported by \citet{Schmidt2021}, clearly indicating instrument-related wavelength-calibration systematics. These could be drastically reduced, to the level of a few $\mps$, by properly taking into account the instrumental LSF \citep{Schmidt2024}.

 Here, we of course as well employ the approach presented in detail by \citet{Schmidt2024}, in which  the observed line is fitted with a model of the intrinsic line shape (here a Gaussian with vanishing width) that is convolved with the non-parametric model of the instrumental LSF derived from the $18\,\mathrm{GHz}$ LFC spectra. Despite this, Figure~\ref{Fig:Accuracy_LFC-BLUVES} shows strong discrepancies between Slice~a and b of Fiber~A. These reach levels of up to \VAL{$\approx15\,\mps$} and correlate directly with the spectral order structure of the echelle spectrograph.
 This is a strong indication that the cause of the discrepancy is related to the spectrograph and not the laser sources themselves.
 We can exclude charge-transfer inefficiencies \citep[e.g.][]{Goudfrooij2006, Bouchy2009, Blake2017, Blackman2020}, since both slices receive by design basically identical flux levels.
 Also, the accuracy of the adopted calibration scheme, using the 18\;GHz LFC for LSF and wavelength calibration, has been validated by \citet{Schmidt2025} against the accurately known features of an iodine absorption cell, finding very good agreement at a level better than \VAL{$\lesssim5\,\mps$}.
 The pattern of the discrepancies found here resembles at least qualitatively the one found in \citet{Schmidt2021} where a Gaussian profile had been assumed for the LSF. This suggests that for the $34\,\mathrm{GHz}$ EOM comb, the effective LSF on the detector is different from the one of the $18\;\mathrm{GHz}$ comb  used to construct the LSF model and the wavelength solution. This, however, is most-likely not related to the intrinsic line shape of the laser, which would affect both slices equally. Instead, we suspect an illumination effect in the fiber train leading to different LSFs in the two slices of Fiber~A which are both not identical with the one from the $18\,\mathrm{GHz}$ comb.

 This is clearly a severe issue. The fundamental assumption of the ESPRESSO calibration strategy (and most other spectrographs) is that all sources are imaged in an identical and comparable way on the detector. Only in this way, it is possible to use one source to calibrate the spectrum of another. Accordingly, much care has been taken in the instrument design to ensure this, for instance by using octagonal fibers \citep{Bouchy2013} and a \textit{double scrambler} \citep{Hunter1992} in the fiber train.
 Despite this, \citet{Schmidt2025} have showed within the context of the iodine absorption cell experiment that the light injection geometry into the spectrograph fibers can lead to significant effects on the spectrum level. However, they found differences of only \VAL{$5$ -- $10\,
\mps$} under highly non-standard illumination conditions.

 Therefore, it is quite surprising that we find here nearly twice the effect while the light from both combs is fed via the same optical fiber from the ESPRESSO calibration unit to the spectrograph front-end and injected there in an identical way.
 The difference in the illumination pattern of the fibers, which ultimately propagates to the detector, must occur at the port-selector units. Here, the facility LFC is connected to input port~4, while our $34\,\mathrm{GHz}$ LFC was using the spare input port~6.
 Issues could in principle also arise earlier in the fiber train, but here we used for our setup the active mode scrambler (see Section~\ref{Sec:ModeScrambler}) based on a rotating diffuser disc, which should ensure a very homogeneous illumination of the fiber leading to the port-selector units.

 Our suspicion is therefore that an optical misalignment in the port-selector units causes an improper illumination of the output fiber when the input ports~6 are selected.
 Since these spare input ports have never been used on a regular basis, their alignment status must be considered undetermined. Unfortunately, the port-selector units are closed boxes and it was impossible to inspect them in detail or perform a re-alignment procedure.

 To make this even more puzzling, the accuracy in Fiber~B is fine, as shown in Figure~\ref{Fig:Accuracy_LFC-BLUVES}, bottom panel. Both slices agree reasonably well, within a \VAL{few $\mps$}, and show only little residual structure correlated with the order structure of ESPRESSO. Also, the absolute difference is close to zero, showing that both combs do indeed provide the same absolute wavelength scale. Some systematics are visible in the bottom panel of Figure~\ref{Fig:Accuracy_LFC-BLUVES}, but these are almost at the usual level found in similar comparisons by \citet{Schmidt2024} and \citet{Schmidt2025}.
 The fact that we find good accuracy for Fiber~B, but large discrepancies for Fiber~A is highly surprising, because the throughput for the spare input in Fiber~B was only about 5\,\% of the one for Fiber~A and moving the carriage in the Fiber~B selector unit to a slightly different position actually improved the throughput. This clearly shows that there was a severe misalignment of the Fiber~B unit.  Nevertheless, we are convinced that also the alignment for Fiber~A was not within nominal parameters and ultimately lead to the significant inaccuracies shown in Figure~\ref{Fig:Accuracy_LFC-BLUVES}.

 The ESPRESSO fiber feed is supposed to be as robust as possible against fiber-injection geometry effects and provide a high degree of scrambling.
 The observations reported here and also the results from \cite{Schmidt2025} clearly indicate that this is not as perfect as desired.
 In \cite{Schmidt2025}, an effect up to \VAL{$5$ -- $10\,\mps$} was observed in the wavelength measurements when the injection geometry deviated significantly from nominal. Also here, we suspect that there must be a large deviation from the optimal alignment. This clearly exaggerates the magnitude of the effect to be expected under nominal conditions, however, no optical system is ever perfectly aligned and it might be possible that at least some part of the remaining systematics of a few $\mps$ reported by \citet{Schmidt2024} and \citet{Schmidt2025} could also be related to slight differences in the fiber illumination between the different input ports. When aiming for further improved accuracy and RV stability, such effects need to be better understood and new solutions developed, i.e. more advanced fiber feeds, which increase the robustness of the spectrograph w.r.t. the light injection geometry. This is relevant for calibration exposures of various light sources, but also for science exposures which naturally occur under varying seeing conditions.

 Nevertheless, it would be highly desirable to further investigate the current issue with the spare input ports of the ESPRESSO calibration unit and determine why it leads to the observed inaccuracies in the wavelength measurements. There are at least two possible tests one could perform:
 The ESPRESSO front-end is equipped with cameras intended for field- and pupil-stabilization during scientific exposures \citep{Megevand2014, Pepe2021}. In principle, they could also be used to image the  near- and far-field of the spectrograph fiber and therefore document the fiber-injection geometry of the light fed from the calibration unit.
 Unfortunately, these cameras are not turned on during calibration exposures. Therefore, no such data is available. In a future experiment, one could activate the cameras manually, acquire dedicated images of the near- and far-field, and thus gain insights about the injection geometry and whether it is different for light coming from the various input ports of the port-selector units.
 It is not entirely clear how informative these frames would be, given that the cameras are just equipped with broadband filters and that the discrepancy between slices in Figure~\ref{Fig:Accuracy_LFC-BLUVES} amounts to \VAL{$15\,\mps$}, which corresponds to just \VAL{0.6\,\%} of the FWHM of the LSF. However, such a test would be entirely non-invasive, relatively easy to carry out, and could at least check whether there is a major irregularity in the fiber illumination.

 Another experiment that should be performed is to feed the same stable calibration source through different input ports of the port-selector units and compare the spectra recorded on the detector. This experiment would require some intervention to the calibration unit. One would have to unplug the currently connected input fibers and plug the desired source into the various input ports.
 Since the ThAr HCL and the FP are used for daily science operations and long-term monitoring campaigns, these calibration sources should better not be touched. However, one could at least connect the $18\,\mathrm{GHz}$ facility LFC to the spare input ports and perform a comparison as shown in Figure~\ref{Fig:Accuracy_LFC-BLUVES}. If one finds for Fiber~A the same discrepancy when using the exactly identical calibration source on port~4 and 6, it would be the ultimate proof that the issue stems indeed from the fiber feed and in particular the port-selector units. Furthermore, it would also make sense to swap the A and B inputs to verify that they provide equivalent wavelength information.
 With these relatively simple experiments, substantial insights into the nature of the problem could be gained.

 We stress that this fiber-illumination issue is not specific to ESPRESSO. The availability of two spectral slices and the in general very high precision and accuracy of the instrument just makes it possible to detect and characterize such effects.  Maybe ESPRESSO has due to the pupil slicer design a more complicated LSF and might be more sensitive to the fiber-injection geometry, but the general issue affects all Extreme-Precision Radial-Velocity (EPRV) spectrographs. In fact, a systematic offset between two LFCs that ultimately could not be resolved was also found during an experiment with HARPS \citep{Probst2016, Milakovic2020a}. It is therefore highly important to gain further insights and equip future spectrographs with an improved fiber feed that adds additional robustness w.r.t. the fiber-injection geometry.

\section{Wavelength Calibration Stability}
\label{Sec:Stability}

\begin{figure*}
 \includegraphics[width=\linewidth]{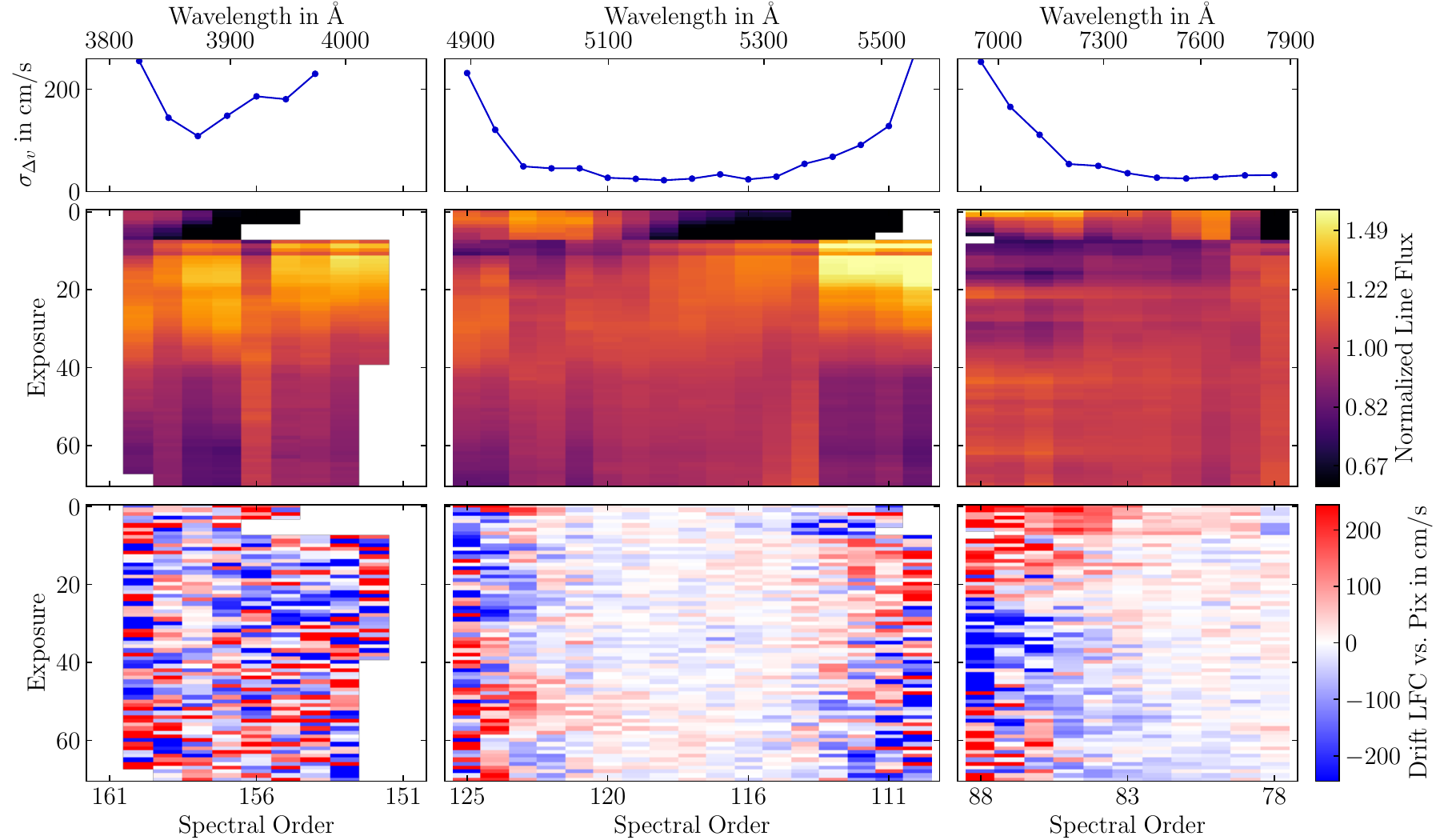}\\%
 \caption{ Visualization of various properties derived from a sequence of \VAL{71} LFC exposures, spanning approximately \VAL{80\,minutes}, obtained on 9 December 2024. The three columns of panels correspond to the spectral ranges covering the fourth, third, and second harmonic of the $34\,\mathrm{GHz}$ EOM comb. The top panels give for each spectral order the mean photon-limited RV precision obtained in one exposure. The second row shows for each spectral order the relative change of the line fluxes along the sequence of exposures, with the time-averaged mean defined to be unity. The bottom panels give the inferred drift of the LFC lines w.r.t. the detector coordinates, averaged over individual spectral orders. All data in the plot were extracted from Slice~a, Fiber~A of the ESPRESSO spectra.}
 \label{Fig:Sequence_2024-12-09_Flux}
\end{figure*}

 Another highly important aspect is of course the \textit{stability} of the derived wavelength calibration. This is fundamentally different from the \textit{accuracy} discussed in the previous section. The substantial systematics presented in Figure~\ref{Fig:Accuracy_LFC-BLUVES} thus do not necessarily have a direct impact on the stability, as long as they do not change with time. For the one week of measurements at Paranal, we have not found any significant change in that pattern. We therefore continue to assess the stability derivable from the $34\,\mathrm{GHz}$ EOM comb spectra.

 For this, we use a sequence of \VAL{71} LFC\,--\,FP exposures taken on 9 December 2024 over a period of slightly more than 80 minutes. Here, our EOM comb was fed to Fiber~A of ESPRESSO and the FP simultaneously to Fiber~B, allowing a precise monitoring of the instrumental drift over the course of the sequence and a direct comparison of the two sources. Here, the FP is best suited as a reference since it is (on short timescales) the most stable of the available calibration sources \citep{Schmidt2025} and covers the full wavelength range.

 The data was reduced in the standard way, using the empirical LSF model inferred from the $34\,\mathrm{GHz}$ EOM comb to fit the lines, and is shown in Figure~\ref{Fig:Sequence_2024-12-09_Flux}. The top panels of the plot show for fourth, third, and second harmonic the mean RV precision achievable in each spectral order of the spectrograph.
 The second row of panels displays the relative change of the flux during the sequence of LFC exposures. Clearly, the flux was not constant but there were variations, the most severe ones during the first approx. \VAL{10~minutes}. After this period, the focus of the chip-injection stage was optimized again and the box containing the spectral broadening setup closed. For the remainder of the sequence, the flux levels showed much better stability, nevertheless with some significant variations.

 While the frequencies of the individual laser lines are highly stabilized and independent of the flux levels, the wavelength information extracted from the LFC spectra might nevertheless be sensitive to it. The main reason for this is the limited spectral resolution (and sampling) of the spectrum compared to the desired wavelength calibration stability. The FWHM of the individual lines recorded on the detector is about $2.3\,\kmps$, more than 10\,000\,$\times$ as wide as the desired RV stability. Even for individual lines, the typical photon-noise limited precision is just 0.1\% of the lines FWHM. Adverse effects related to flux changes can therefore easily cause significant systematics, e.g. by skewing the observed profile, not of the line itself but the underlying spectrally-diffuse background. Here, one also has to consider that the data processing itself is not perfect but inherently introduces systematics, e.g. due to the non-optimal spectral extraction \citep[see e.g.][]{Bolton2010, Zechmeister2014, Schmidt2021} or when an incorrect instrumental LSF is adopted while fitting the lines \citep{Schmidt2024}. Instabilities in flux can therefore cross-couple and lead to instabilities in the derived wavelength information.
 The FP calibration source is in this regard very easy and forgiving, because fitting its marginally resolved lines (intrinsic width $\approx2\,\mathrm{GHz}$ or $\approx1\,\kmps$ at $500\,\mathrm{nm}$) is not particularly sensitive to the adopted instrumental LSF and its flux is stable to better than a per-cent over hours and days. If one is just interested in relative drifts, it is often even sufficient to rely on the simple but very efficient \textit{gradient-method} \citep{Bouchy2001}.
 LFCs are quite the opposite. Their extremely narrow lines impose high demands on the accuracy of the adopted instrumental LSF and the necessary non-linear effects in the spectral broadening process make it very difficult to deliver spectra with truly stable flux levels.

\begin{figure*}
 \includegraphics[width=\linewidth]{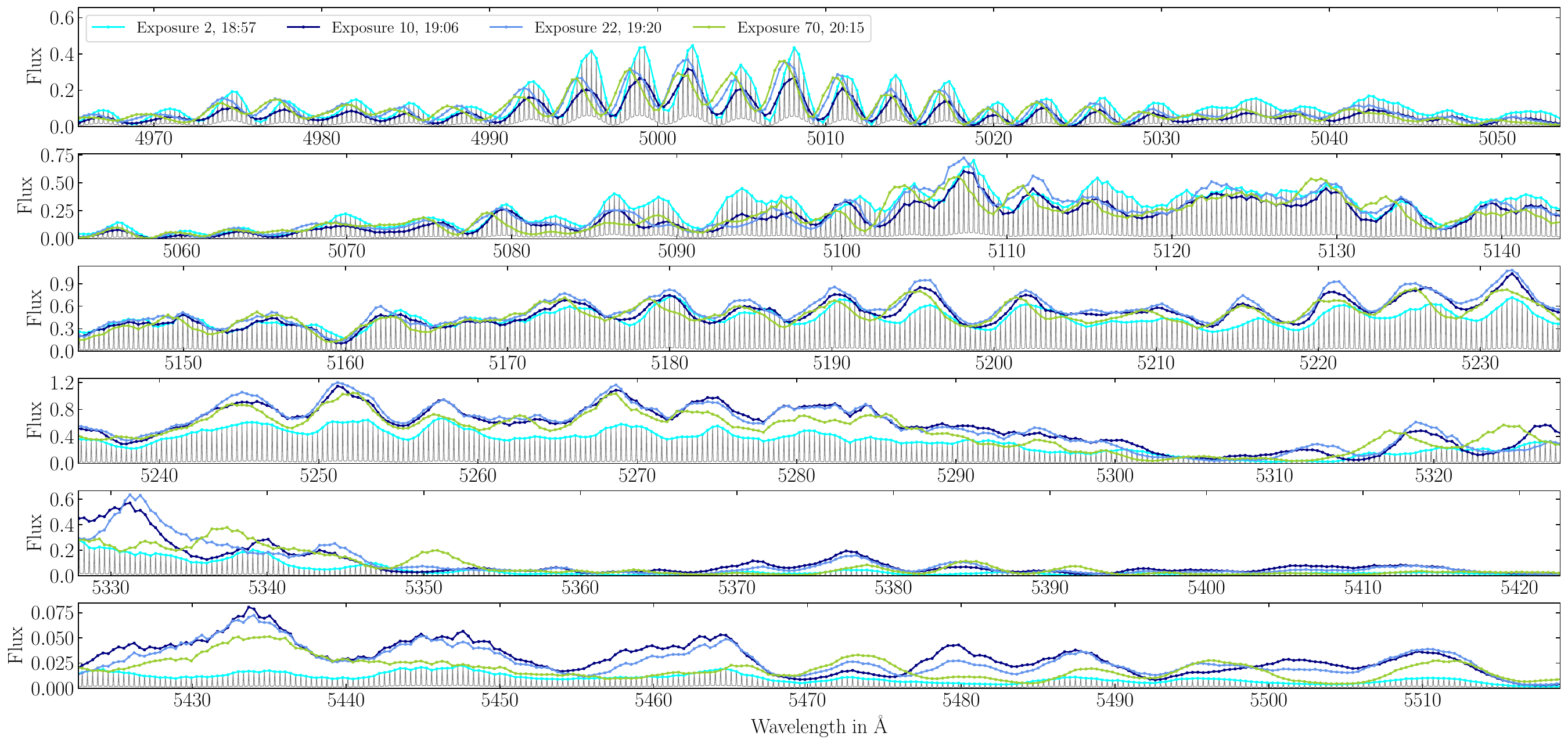}\\%
 \caption{ Detailed visualization of the flux variation of the LFC spectrum for four selected exposures of the sequence presented in Figure~\ref{Fig:Sequence_2024-12-09_Flux}. For Exposure~\VAL{2 (18:57)}, the full spectrum of the third harmonic is shown. In addition, the flux envelope based on the peak heights of the lines is indicated. For three other exposures (\VAL{Exposure 10 (19:06)}, \VAL{22, (19:20)}, and \VAL{70 (20:15)}, times given in UTC), only the flux envelope is shown. }
 \label{Fig:Sequence_2024-12-09_PhaseMatching}
\end{figure*}

 For the analyzed sequence of LFC frames, Figure~\ref{Fig:Sequence_2024-12-09_Flux} reveals only the order-averaged flux changes. The comb, however, exhibits in addition to this a substantial substructure that originates from the varying phase-matching conditions along the TFLN waveguide. This substructure also changes with time, without necessarily causing a net flux change.
 To visualize this effect, Figure~\ref{Fig:Sequence_2024-12-09_PhaseMatching} shows for one exposure the full spectrum of the \VAL{third} harmonic and for \VAL{three} other exposures of the same sequence just the flux envelope. It becomes clear that the dominant effect is not necessarily some global or at least large-scale decrease or increase in flux, but variations on much smaller scales, sometime only affecting ten lines or less. We notice that this substructure and its variability become stronger in the wings of the harmonics, and also with increasing harmonic number.
 In particular when this highly structured pattern moves in wavelength, certain lines gain in flux while others might lose tremendously. The typically adopted strategy for inferring a drift is to first compute for each line the drift between exposures and then average over the lines. For a specific line, the uncertainty in the drift estimate is dominated by the exposure in which the line has the lowest flux. If the structure of the flux envelope changes strongly, a good part of the in-principle available S/N can therefore not be exploited.
 Also, comparing the same line at different flux levels might lead to detrimental effects. For instance, a line at a valley of the flux envelope might be affected more strongly by the background than one on a ridge, which itself can have an impact on the determined line centroid if the background is not modeled perfectly or an inaccurate LSF model is used.
 
 The goal must therefore be to produce a flat spectrum with as little structure as possible that in addition also remains mostly stable. 
 Unfortunately, fully controlling the phase-matching in the waveguide is hardly possible and the desire for a flat spectrum might be in conflict with other design goals. Certain waveguide designs we tested did actually produce a brighter or broader spectrum, but at the expense of extreme substructure. Since this has a negative impact on the quality of the inferred wavelength information, the waveguide used for the spectra shown in Figure~\ref{Fig:Sequence_2024-12-09_PhaseMatching} was considered the better compromise.

 A fully complementary approach is therefore to make the data reduction as robust as possible to unavoidable flux changes. For this, an accurate modeling of the instrumental LSF has proven to be essential.
 Despite the much larger computational effort, we therefore fit all LFC lines in the sequence with our empirical LSF model.
 Here, we use the LSF inferred directly from the $34\,\mathrm{GHz}$ EOM comb, however, from an exposure taken on \VAL{10 December 2025} that is not part of the sequence itself to avoid circularity. For third and second harmonic, we verify that using the LSF derived from the $18\,\mathrm{GHz}$ comb provides virtually identical results.
 The bottom panels of Figure~\ref{Fig:Sequence_2024-12-09_Flux} show the change of the inferred line position w.r.t. the detector coordinates.
 Using the accurate LSF model in the fitting process, these drift measurements appear rather clean and consistent. Some instrumental drifts are present, which most-likely originate from a \textit{breathing} of the detectors and some other drifts. Importantly, there is no strong correlation with the flux changes and we confirm that both spectral slices provide results consistent with each other (see detailed analysis below).

 \begin{figure}
 \includegraphics[width=\linewidth]{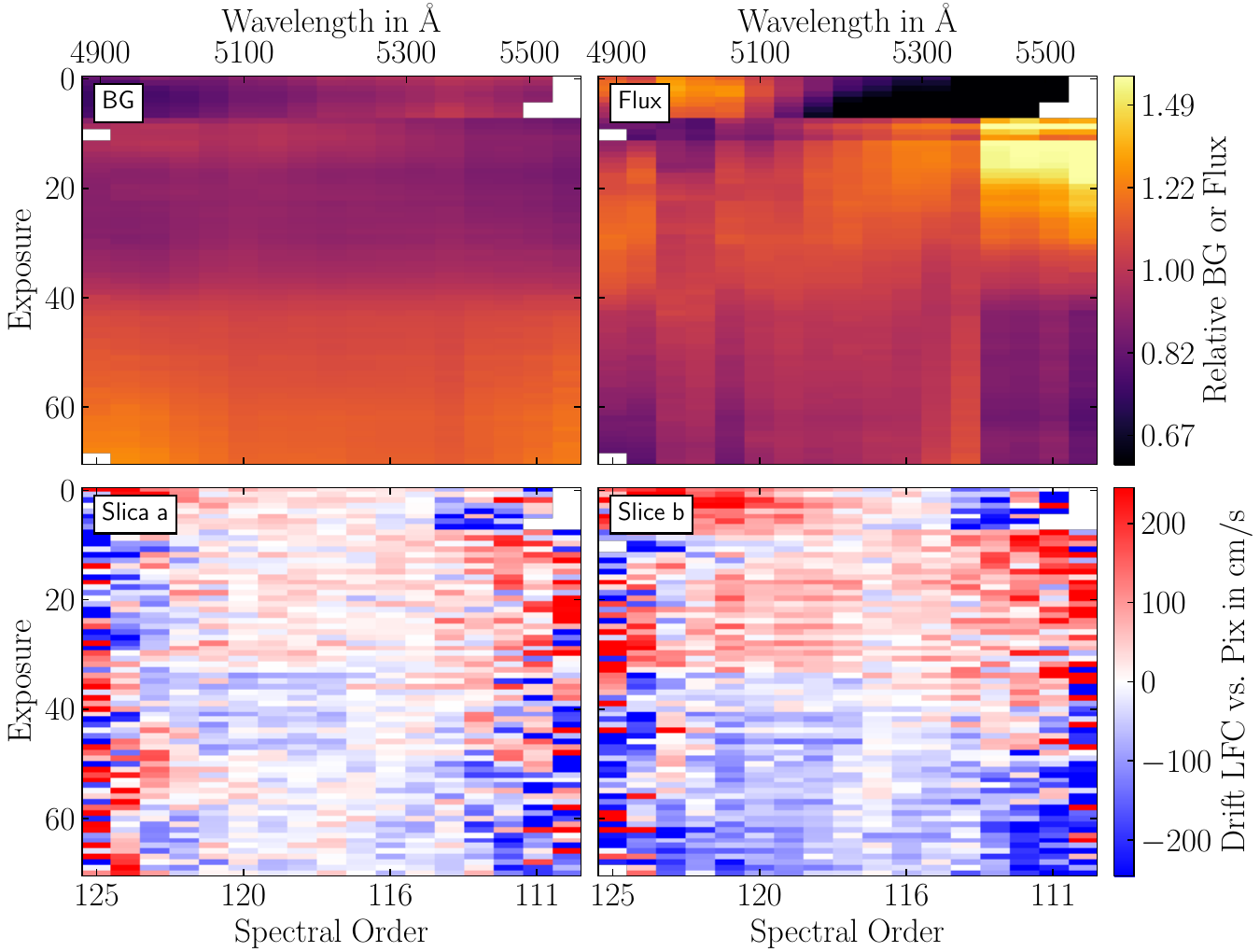}\
 \caption{ Comparison of the drift inferred from Slice~a and b when assuming Gaussian line profiles. Based on the same raw data as Figure~\ref{Fig:Sequence_2024-12-09_Flux}. The two top panels visualize for each spectral order the relative change of the background fraction (left) and the line flux (right) along the sequence of exposure. w.r.t. the temporal mean The bottom panels show the inferred drift of the LFC lines w.r.t. the detector coordinates for Slice~a (left) and Slice~b (right). }
 \label{Fig:Sequence_2024-12-09_Gaussian}
\end{figure}

 This, however, is different when using a Gaussian LSF. In Figure~\ref{Fig:Sequence_2024-12-09_Gaussian}, we show for the third harmonic the drifts inferred from both slices side-by-side, assuming a Gaussian LSF. Clearly, both drifts measurements are not consistent anymore. While we find approximately zero drift in Slice~a, Slice~b exhibits a change of roughly $-2\,\mps$ over the course of the sequence. This is unphysical, since both slices are recorded next to each other on the detector and should be subject to the identical drift. In addition, one notices, in particular for Slice~b, a correlation between the inferred drift and changes in the flux level of the lines, but also an anti-correlation with the background fraction. 
 Here, the top-left panel of Figure~\ref{Fig:Sequence_2024-12-09_Gaussian} visualizes the relative change of the background fraction, revealing variations by roughly $\pm$30\,\% relative to the temporal mean. 
 Changes in flux and background are identical for both slices, but obviously, they affect the drift measurements in Slice~b more strongly. The reason for this is most-likely that the two slices have significantly different LSFs \citep[see][]{Schmidt2024}.
 We stress that this effect is not unique to out $34\,\mathrm{GHz}$ EOM comb. Similar issues have also been noticed for the $18\,\mathrm{GHz}$ facility LFC.
 One therefore has to conclude that for LFCs which provide \textit{complicated} spectra, i.e. with significant and variable background, substructure in the flux envelope, and unavoidable fluctuations in the flux levels, it is of crucial importance to fit the lines using an accurately model of the instrumental LSF and the spectrally-diffuse background. Only by this can the inherent frequency stability of the LFC be translated into good stability of the instruments wavelength calibration.

\begin{figure}
 \includegraphics[width=\linewidth]{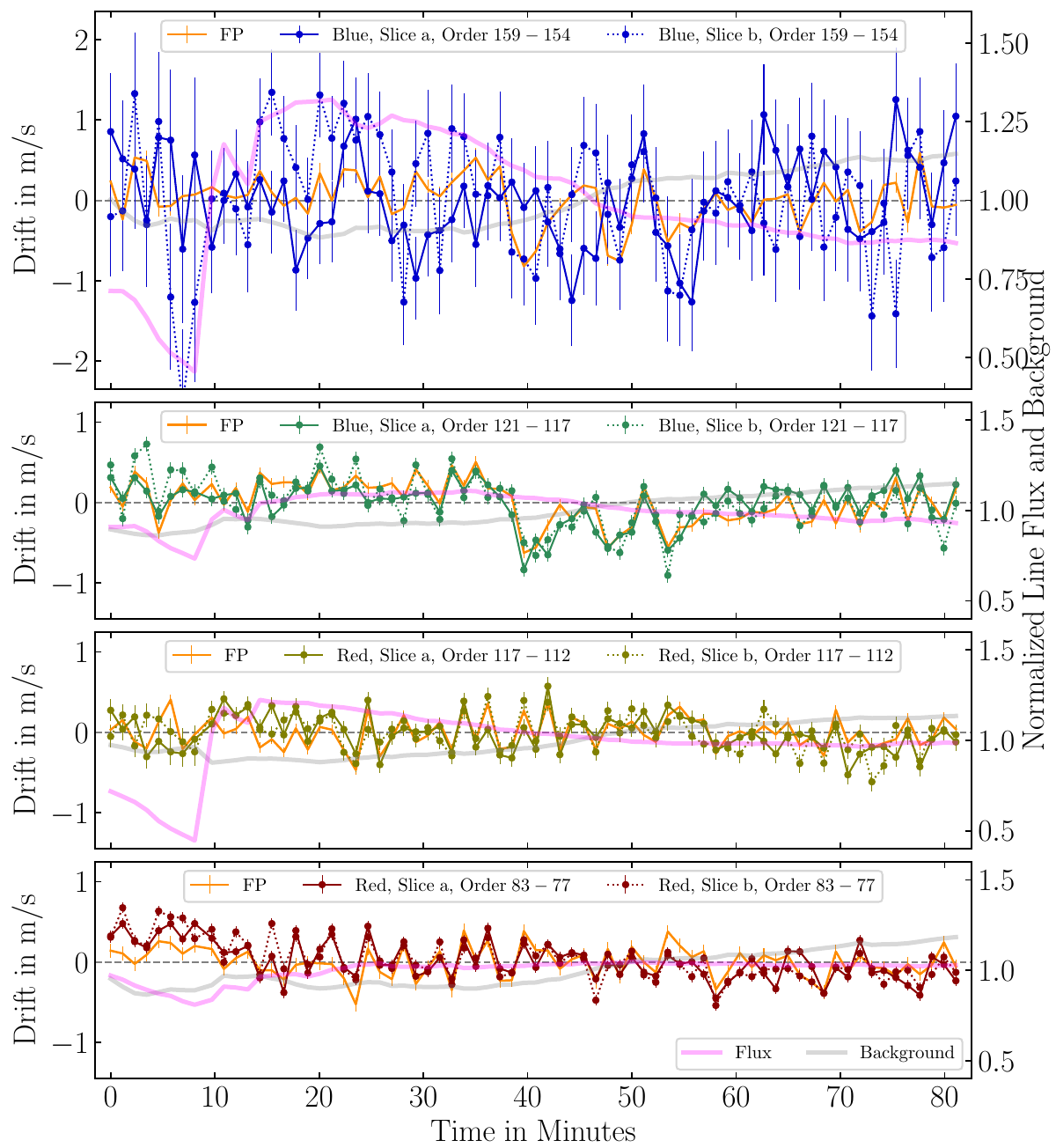}\\%
 \caption{ Inferred drift of the spectral features in the FP and LFC spectra, given in terms of velocity and relative to the detector coordinates, over the course of the sequence presented Figure~\ref{Fig:Sequence_2024-12-09_Flux}. The four panels show from top to bottom the drift averaged over different spectral regions, i.e. the fourth harmonic, the part of the third harmonic recorded by the blue arm of the instrument, the third harmonic from the red arm, and the second harmonic. The exact spectral orders are stated. Each panel shows the drift inferred from Slice~a and Slice~b of the LFC spectrum, as well as from the FP. In addition, the relative change of line flux and background fraction, averaged over the corresponding spectral region is given.}
 \label{Fig:Sequence_2024-12-09_Drift}
\end{figure}

To perform a more quantitative assessment of the drifts shown in Figure~\ref{Fig:Sequence_2024-12-09_Flux}, which do make use of such a careful modeling of the LSF, we spectrally average these measurements from individual orders. However, we still keep the different harmonics and the two slices separate. In addition, we also split the third harmonic in two parts to embrace that this part of the spectrum is recorded separately by the blue and red arms of ESPRESSO, which overlap in spectral order 117.
The result of this is shown in Figure~\ref{Fig:Sequence_2024-12-09_Drift}, together with the drift inferred from the simultaneous FP data recorded in Fiber~B. In the FP spectrum, lines were fitted with Gaussians and both slices combined, because there are no difficulties that would require a more complicated approach.

As can be seen in the plot, the drifts from both slices show a remarkable agreement, with each other but also with the drift inferred from the FP. Some discrepancies are noticeable for the first \VAL{10} minutes where the flux level were different and unstable, but the agreement for the remainder of the sequence is excellent.
One also notices that the pattern of the measured wavelength shifts is similar for fourth and third harmonic recorded by the blue arm of the spectrograph (clearly visible for the FP, but limited by the S/N in the fourth harmonic of the comb) and also for third and second harmonic from the red one. This shows that the observed shifts of the lines originate indeed from the instrument, most-likely the detector assemblies, and are not inherent to the calibration sources. The LFC recorded in either of the slices, as well as the FP, would all be well-capable to track the drift of the instrument with great precision.
Taking the difference between the drift inferred from the LFC and the FP, we find a RMS scatter of the residuals of about \VAL{$17\,\cmps$} for any of the six curves shown in Figure~\ref{Fig:Sequence_2024-12-09_Drift} that correspond to third or second harmonic. For these, the combined photon noise from LFC and FP spectra amounts to \VAL{$10-12\,\cmps$}. Thus, despite the changes in background and line flux in the LFC spectra (also indicated in Figure~\ref{Fig:Sequence_2024-12-09_Drift}), does the actual scatter not exceed the propagated noise by more than \VAL{50\,\%}.
For the fourth harmonic, the photon-noise limited precision of each drift measurement is substantially worse, approximately \VAL{$60\,\cmps$}.
The reason for this are the lower flux of the LFC, the inferior throughput of the instrument, and the photon-counting nature of the detectors.
The RMS scatter between LFC and FP, however, is fully consistent with the propagated photon noise and found to be \VAL{$\approx60\,\cmps$} as well.
This demonstrates that despite the complications inherent to the LFC spectrum (structured and variable flux envelope, intensity changes, diffuse background, ... ) and the inaccuracies in the wavelength calibration caused by the ESPRESSO fiber feed injection (see Figure~\ref{Fig:Accuracy_LFC-BLUVES}), highly stable wavelength information can be extracted from the spectrum of the $34\,\mathrm{GHz}$ EOM comb. This, however, crucially relies on a sophisticated and accurate, but also complex and computationally expensive, characterization and forward-modeling of the non-Gaussian and in some parts strongly skewed instrumental LSF of ESPRESSO. This test therefore demonstrates the general feasibility of using LFCs for wavelength calibration of astronomical spectrographs, but also the challenges of exploiting their full potential and dealing with the aspects that make their spectra more difficult to analyze than for instance FPs.

\section{Summary and Conclusions}

Here, we have presented an updated version of the LFC setup described in \citet{Ludwig2024} and performed a detailed analysis of its spectral properties based on extensive tests with the ESPRESSO high-resolution spectrograph at the Paranal observatory.

Major differences compared to our previous test are the substantially increased repetition rate of the fundamental \VAL{$1.56\,\mathrm{\mu{}m}$} EOM comb, now operating at $34\,\mathrm{GHz}$, and the introduction of a high-finesse FP cavity (\VAL{$\mathcal{F}\approx725$}) for noise suppression.
Spectral broadening and harmonic generation was again facilitated by custom-manufactured TFLN waveguides, based on the design concepts presented in \citet{Ayhan2025}.
A vast improvement in our analysis of the LFC spectra comes from the superior capabilities of ESPRESSO, which provides larger spectral coverage, particularly high resolution ($R\approx135\,000$), exquisite stability, an accurately characterized LSF, and a mode-locked $18\,\mathrm{GHz}$ facility LFC, allowing to directly compare these two vastly different LFCs and to perform a detailed characterization of our LFC spectra.

In Section~\ref{Sec:DiffuseBackground}, we presented a thorough analysis of the spectrally-diffuse background in the LFC spectra. For this, we operated the $34\,\mathrm{GHz}$ comb without any special noise filtering, with a tailored mask imposed by the waveshaper, and in the nominal configuration with the high-finesse FP cavity. As shown in Figure~\ref{Fig:BackgroundFraction}, these measures lead to a dramatic improvement of the background fraction (by about two orders of magnitude) and a vast increase of the flux contained in the lines. Nevertheless, and also due to the large repetition rate, is the background fraction still relatively high and exhibits the known pattern of approximately quadratic increase with mode number or equivalently with spectral broadening away from the center of the harmonics.

In addition, we performed a detailed analysis of the LFC line width (see Section~\ref{Sec:LineWidth}). Here, ESPRESSO has, despite its high resolution, an instrumental line width of \VAL{$\approx2.2\,\kmps$}, corresponding to \VAL{$\gtrsim4\,\mathrm{GHz}$} and thus far in excess of the precision that can be achieved with heterodyne beat note measurements. Nevertheless, by comparing the apparent line width to the one observed in the mode-locked LFC, 
we confirm that also our EOM comb delivers truly unresolved lines in the centers of the third and second harmonic. Towards the edges of the harmonics, however, the lines become noticeably wider (Figure~\ref{Fig:LineWidth_Harmonics}), as expected from the theoretical description presented in \citet{Ludwig2024}.
For the fourth harmonic, in the blue to near-UV spectral range without coverage by the mode-locked LFC, no such detailed assessment is possible, but a comparison to the thorium lines still indicates that the lines of the EOM comb also here have no noticeable intrinsic width.

LFCs are also currently the only type of calibration sources that facilitates an accurate characterization of the instrumental LSF \citep[e.g.][]{Ninan2019, Zhao2021, Schmidt2024, Sekhar2024}. Here, the line spacing of the comb limits the width over which the LSF can be observed. Leveraging on the large repetition rate of the $34\,\mathrm{GHz}$ EOM comb, we characterize the extended wings of the ESPRESSO LSF and find structures not visible in the $18\,\mathrm{GHz}$ LFC spectra (Figure~\ref{Fig:LSF_LFC-BLUVES}). We demonstrate that these are real and provide a better model to the thorium lines (Figure~\ref{Fig:LSF_chi2}), however, we also stress that we did not find any difference for the actual wavelength measurements, since the majority of the information content is anyway located in the core of the lines.

Regarding wavelength calibration accuracy (Section~\ref{Sec:Accuracy}), we found a very surprising and concerning result. The comparison of the wavelength information inferred from the two combs in Figure~\ref{Fig:Accuracy_LFC-BLUVES} shows, despite a careful modeling of the LSF,  in Fiber~A local discrepancies up to \VAL{$15\,\mps$}, which highly correlate with the echellogram structure of the spectrograph. This indicates that the two LFCs produce different instrumental LSF on the detector.
We are convinced, however, that this is actually no issue with the lasers themselves, but relates to the fiber feed and in particular to the spare inputs of the port selector units.
We suspect that the inputs used for the EOM comb are misaligned and lead to an improper light-injection into the calibration fiber, which propagates to the spectrograph and there creates a LSF on the detector that is different from the one inferred from the $18\,\mathrm{GHz}$ comb, fed via another set of input ports.
This is a clear issue and invalidates the fundamental calibration concept, which requires that all sources (science or calibration) are imaged onto the detector in an identical and comparable way.
The misalignment for these input ports might be particular severe and far out of spec, but no optical system is perfectly aligned and the discovered discrepancies indicate that in general a higher level of robustness of the fiber feed w.r.t. the light injection geometry is needed to ensure a better level of accuracy and stability.
The observed inconsistency in Fiber~B is reasonably good though, typically a \VAL{few $\mps$} between slices measured in small blocks and \VAL{almost zero} when comparing both combs and averaging over the full wavelength range. 

Despite these issues regarding the obtained wavelength calibration accuracy, we achieve good results in terms of stability (see Section~\ref{Sec:Stability}). Comparing the two spectral slices to each other as well as to the simultaneous FP in Fiber~B, we observe excellent consistency over a sequence of \VAL{80\,min}.
For the rms scatter between LFC and FP drift in any of the slices of third or second harmonic, we find about \VAL{$17\,\cmps$}, given a photon noise of \VAL{$\approx11\,\cmps$}. In the fourth harmonic, drifts are fully consistent with the noise of \VAL{$60\,\cmps$}, as shown in Figure~\ref{Fig:Sequence_2024-12-09_Drift}.
However, this stability of the wavelength calibration can only be achieved by applying to all LFC lines the full forward-modeling technique based around the non-parametric description of the instrumental LSF \citep{Schmidt2024}.
Assuming e.g. Gaussian line shapes, results in apparent differential drifts between the spectrograph slices in excess of \VAL{$2\,\mps$}, which are unphysical and incorrect (Figure~\ref{Fig:Sequence_2024-12-09_Gaussian}). The reason for this are unavoidable variations in the LFC line and background fluxes, combined with imperfections in the data treatment, e.g. non-perfect spectral extraction and in particular an inaccurate LSF.
This highlights that LFCs bear great potential for the wavelength calibration of astronomical spectrographs, but also provide spectra that are difficult to analyze and demand very careful and accurate methods for data reduction and calibration to actually unleash their full potential.

Future developments should therefore aim at a further reduction of the spectrally-diffuse background flux, a less structured spectrum, and better flux stability.
However, due to the non-linear processes required for frequency conversion, LFC spectra might never exhibit the outstanding flux stability of e.g. FP etalons.
Usage of LFCs for wavelength calibration therefore mandates the development of sophisticated data reduction pipelines that can deal with
variations in the line fluxes, spectral envelopes, and background.
It is also necessary to implement a proper characterization and forward-modeling of the instrumental LSF to adequately deal with the extremely narrow LFC lines. This then has to be applied to all spectra, calibration and science. Although complex and computationally expensive, it then allows to fully benefit from the outstanding frequency accuracy and stability of the LFC lines.

\section*{Acknowledgements}

We thank the staff of the European Southern Observatory (ESO) for their support and assistance at the Paranal site.
This project was funded by the SNF synergia grant CRSII5-193689 (BLUVES).
This work has been carried out within the framework of the National Centre of Competence in Research PlanetS supported by the Swiss National Science Foundation.
This project has received funding from the European Research Council (ERC) under the EU's Horizon 2020 research and innovation program (grant agreement No 853564) and through the Helmholtz Young Investigators Group VH-NG-1404.
This research has made use of Astropy, a community-developed core Python package for Astronomy \citep{Astropy2013,Astropy2018,Astropy2022}, and Matplotlib \citep{Hunter2007}.

\section*{Data Availability}


All data used for this study were obtained under ESO program ID~114.28HD.001 and was publicly available from the ESO archive facility \url{http://archive.eso.org/cms/data-portal.html} but has vanished from there. 
Pipeline-reduced data is available from the DACE platform \url{https://dace.unige.ch/observationSearch/} and raw data can be shared upon personal request.


\bibliographystyle{mnras}
\bibliography{Literature}


%
%


\label{lastpage}
\end{document}